\documentclass[twocolumn,twocolappendix,numberedappendix]{openjournal}

\usepackage{amsmath}

\usepackage{xcolor}
\usepackage{textgreek}
\usepackage[utf8]{inputenc}
\usepackage[english]{babel}

\usepackage{hyperref}
\hypersetup{
    unicode, 
    colorlinks=true,
    linkcolor=linkcolor,
    citecolor=linkcolor,
    filecolor=linkcolor,
    urlcolor=linkcolor,
}
\usepackage{color,colortbl}
\definecolor{linkcolor}{rgb}{0.0,0.3,0.5}
\usepackage{tensind}
\tensordelimiter{?}
\DeclareGraphicsExtensions{.bmp,.png,.jpg,.pdf}
\usepackage{verbatim}
\usepackage[normalem]{ulem}
\usepackage{orcidlink}
\usepackage{soul}

\graphicspath{ {./} }

\def\d{\mathrm{d}}

\newcommand{\lrb}[1]{\left({#1}\right)}
\newcommand{\lrsb}[1]{\left[{#1}\right]}
\newcommand{\lara}[1]{\left\langle{#1}\right\rangle}
\newcommand{\abs}[1]{\left|#1\right|}
\newcommand{\MBH}{M_{\text{BH}}}
\newcommand{\bs}[1]{\boldsymbol{#1}}

\begin{document}
\title{Formation of Close Binaries through Massive Black Hole Perturbations and Chaotic Tides}

\author{Howard Hao-Tse Huang\orcidlink{0000-0003-4647-2591}}
\email{haotse813@berkeley.edu}
\affiliation{Department of Astronomy, University of California at Berkeley, Berkeley, CA 94720, USA}

\author{Wenbin Lu\orcidlink{0000-0002-1568-7461}}
\email{wenbinlu@berkeley.edu}
\affiliation{Department of Astronomy, University of California at Berkeley, Berkeley, CA 94720, USA}

\begin{abstract}
Hills breakup of binary systems allows massive black holes (MBH) to produce hyper-velocity stars (HVSs) and tightly bound stars.
The long timescale of orbital relaxation means that binaries must spend numerous orbits around the MBH before they are tidally broken apart.
Repeated MBH tidal perturbations over multiple pericenter passages can perturb the binary inner orbit to high eccentricities, leading to strong tidal interactions between the stars.
In this work, we develop a physical model of the MBH-binary system, taking into account outer orbital relaxation, MBH tidal perturbations, and tidal interactions between the binaries in the form of dynamical tides.
We show that when the inner orbit reaches high eccentricities such that the pericenter radius is only a few times stellar radii ($R_*$), the stellar oscillation modes can grow chaotically and rapidly harden the binaries to semi-major axes $a_b\lesssim 10\,R_*$.
We find that a significant fraction (up to 50\%) of initially wide binaries that are in the empty loss-cone regime ($a_b\sim 1.0\,{\rm AU}$) do not undergo Hills breakup as wide binaries, but instead experience chaotic growth of tides and become close binaries.
These tidally hardened binaries provide a new channel for the production of the fastest HVSs, and are connected to other nuclear transients such as repeating partial tidal disruption events and quasi-periodic eruptions.
\end{abstract}


\maketitle

\section{Introduction}

The well-known outcome from the tidal breakup of a binary system by a massive black hole (MBH) is that one star (with positive energy) is ejected to infinity as a hyper-velocity star \citep[HVS;][]{brown15_HVS} and that the other one (with negative energy) is left bound to the MBH. The original work of \citet{Hills1988} and many subsequent works \citep[e.g.,][]{yu03_tidal_break-up, bromley06_HVS_v_distribution, sari10_HVS, rossi14_HVS_velocity_distribution, generozov20_S_stars, yu24_tidal_break-up} only considered the idealized case that a binary remains unperturbed before the tidal breakup, which is only true if the breakup occurs in one outer orbit. However, in reality, the binary's outer orbit undergoes a large number of pericenter passages\footnote{A binary system that is initially on a parabolic orbit may also undergo multiple pericenter passages due to the energy exchange between inner and outer orbits \citep[see][]{Sersante2025}.} (driven by relaxation) before the final breakup and, during this time, the binary's inner orbit can be significantly modified by the MBH's tidal perturbations combined with tidal interactions between the two stars.

Recently, \citet{Bradnick2017} considered multiple pericenter passages along with relaxation of the outer orbit. They found that the inner eccentricity can be excited by the cumulative perturbations of the MBH's tidal forces, and they concluded that most binaries undergo collisions instead of tidal breakups. \citet{Stephan2016,stephan19_eLK} considered binary evolution over multiple outer orbits in a wide range of outer eccentricities using the eccentric von Zeipel-Lidov-Kozai (eZLK) framework \citep{vonZeipel1910,Lidov1962,Kozai1962,Naoz2016}, and found that the eZLK mechanism can induce high inner eccentricities and lead to stellar collisions/mergers.

These two works mentioned above assumed that the
the tidal circularization of the inner orbit is due to damping of the time-dependent equilibrium tide \citep[e.g.,][]{zahn77_tidal_friction, hut81_tidal_friction, Eggleton1998}. In reality, at sufficiently high inner eccentricities (e.g., when the inner pericenter is less than a few stellar radii), we expect the encounters between the two stars to excite strong dynamical tides. Moreover, if the tidally excited modes lead to significant change in the inner orbital period, the pericenter passages may occur at random oscillation phases such that the system is in the chaotic regime \citep{Kochanek1992, Mardling1995}. This leads to rapid growth of the mode amplitude, and the rapid shrinkage of the inner semi-major axis causes the inner orbit to decouple from the outer orbit and hence a violent collision can be avoided. Such chaotic tides have been considered in the orbital migration of hot Jupiter planets in close orbits around host stars \citep{Ivanov2004, Wu2018, Vick2018}. Similarly, we expect binary stars that have undergone chaotic tides to evolve into very close inner orbits.

Tidal breakup of close binaries will deliver the bound stars extremely close to the MBH. Although the fate of these bound stars from Hills breakup are uncertain \citep{Sari2019,Lu2021,Linial2023}, they may be responsible for quasi-periodic eruptions (QPEs) and repeating partial tidal disruption events (TDEs) \citep[e.g.,][]{Miniutti2019,Giustini2020,Arcodia2021,Payne2021, Cufari2022, lu23_QPE_from_EMRI, linial23_QPE_EMRI_TDE,yao25_tidal_heating, Makrygianni2025, pasham24_QPE1_long_term, chakraborty25_QPE_AT2022upj}. A significant fraction of close binaries may also undergo double tidal disruptions if the pericenter radius of the outer orbit is smaller than the tidal disruption radii of both stars \citep{mandel15_dTDEs}.

In this paper, we study the long-term evolution of binary stars including the effects of outer orbital relaxation, tidal perturbations of the inner orbit by the MBH, and dynamical tides excited by the mutual tidal interactions between the two stars. We quantify the fractional outcomes of tidal breakup, collisions/mergers, and chaotic tides. Our main findings are: (1) the repeated MBH perturbations on the binary during the stochastic outer orbit relaxation can drive the binary's inner orbit to high eccentricities, and (2) a significant population of wide binaries, instead of undergoing Hills breakup or collisions, may experience the chaotic growth of tides and become close binaries. Near the completion of this paper, we notice another complementary work by \cite{dodici25_chaotic_tides_KL_and_flyby} who proposed that vector resonant relaxation and perturbations of the inner eccentricity due to fly-bys of field stars can cause the eZLK mechanism to become more effective in bringing the binary to the regime of chaotic tides. There are, therefore, multiple channels of making close binaries via chaotic tides.

This paper is organized as follows. In Section~\ref{sec:Methods} we introduce our physical model of the MBH-binary system. Our results of individual systems and population study are presented in Section~\ref{sec:Results}. We discuss the limitation and implications of our model in Section~\ref{sec:Discussion}, and conclude our findings in Section~\ref{sec:Conclusions}.

\section{Methods}\label{sec:Methods}

\begin{figure*}
    \centering
    \includegraphics[width=\textwidth]{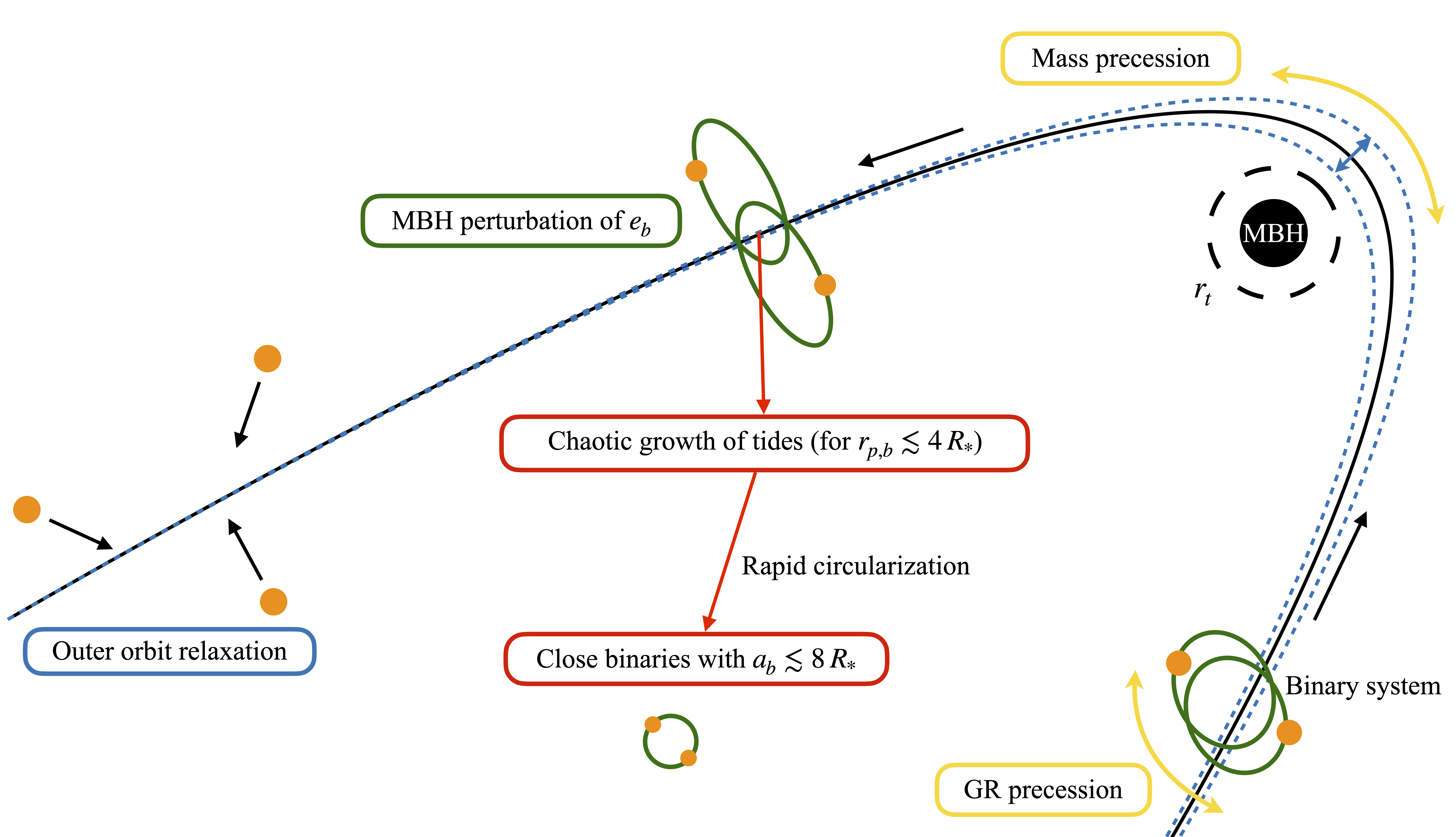}
    \caption{Overview of the physical processes in the MBH-binary three-body system. The binary system is on a highly eccentric orbit around the MBH. On a timescale of $t_{L,\rm relax}$ (eq.~\ref{eq:t-relax-AM}), the outer pericenter radius $r_p$ gradually changes because of the gravitational encounter with other field stars. In our simulation, the gravitational encounters are not resolved; instead, the outer orbit relaxation is handled using a simplified version of Fokker-Planck formulation (Section~\ref{sec:Methods-outer-relaxation}). Every time the binary passes through the pericenter around the MBH, its inner eccentricity is perturbed by the MBH's tidal forces (Section~\ref{sec:Methods-MBH-perturbation}). When the inner eccentricity $e_b$ grows to a sufficiently large value, two binary stars start to interact tidally near the inner pericenter, which causes the energy transfer between the inner orbit and the stellar oscillations. Under certain conditions (eq.~\ref{eq:chaotic-tides-phase-change}), the stellar mode amplitude will grow chaotically,  leading to the rapid hardening of the binary system (Section~\ref{sec:Methods-chaotic-tides}). Finally, there exists various precessions that can affect the MBH perturbations on $e_b$ (Appendix~\ref{sec:App-precessions}).
    }
    \label{fig:system-overview}
\end{figure*}

We consider a binary system on a highly eccentric Keplerian orbit around a MBH. Throughout this paper, we take our own Galactic Center as an example, with MBH mass $\MBH=4.26\times10^6\,M_\odot$ \citep{GRAVITY2020}. The binary system consists of two zero-age-main-sequence (ZAMS) stars of mass $M_*=0.5\,M_\odot$ and radius $R_*=0.478R_\odot$. Our framework is general and can be applied to other MBH and stellar masses as well. In this hierarchical three-body system, we refer to the orbit between the two stars as the inner orbit, and the orbit of the binary system around the MBH as the outer orbit. In this paper, we denote the orbit elements of the inner orbit with an additional subscript $_b$ to distinguish them from those of the outer orbit. When necessary, $_{*,1}$ and $_{*,2}$ are used to emphasize the properties of the first and second stars in the system. The evolution of the binary system is governed by three pieces of physics: (1) the outer orbit relaxation by the gravitational encounters with other field stars, (2) the inner orbit perturbation by the MBH tides during the pericenter passage of the outer orbit, and (3) the influence of the stellar tides on dynamics of the inner orbits. The interaction between MBH and the binary is further affected by the various precessions. The details of these precessions are described separately in Appendix~\ref{sec:App-precessions}.

Figure~\ref{fig:system-overview} provides an overview of the physics in the system. 
On a long timescale over many outer orbits, the binary experiences gravitational interactions with other stars and its outer orbit gradually relaxes. As a result, the outer pericenter $r_p$ may decrease for some systems. The MBH tides will overwhelm the inner orbit and lead to Hills breakup if $r_p$ is reduced to a few times the tidal breakup radius \citep{Hills1988}:
\begin{equation}
    r_t = a_b\lrb{\frac{\MBH}{M_b}}^{1/3},
\end{equation}
where $a_b$ is the semi-major axis (SMA) of the inner orbit and $M_b=M_{*,1}+M_{*,2}=2\,M_*$ is the total mass of the binary system.

However, before $r_p$ reaches $\text{a few}\times r_t$, the perturbation of MBH tides on the inner orbit will already accumulate over several outer orbits and cause a change in the inner eccentricity $e_b$\footnote{For $r_p\lesssim 10\,r_t$, $a_b$ can also be perturbed due to the chaotic three-body interaction (see Section~\ref{sec:Methods-MBH-perturbation}).}.
When $e_b$ is driven to very large values, strong tidal interactions between two binary stars can happen, and in the most extreme case the two stars will collide.
If the inner pericenter radius $r_{p,b}$ is only a few times $R_*$, the tidal forces between the two stars will excite the stellar oscillation modes and transfer the orbital energy to the stellar oscillations \citep{Press1977,Kochanek1992}.
With sufficiently small $r_{p,b}$ and large energy kicks, the amplitudes of the stellar oscillation can grow chaotically and lead to a rapid shrinkage of the inner orbit within one outer orbit \citep{Wu2018,Vick2018}.
The production of hardened binaries has several interesting implications, which are discussed in Section~\ref{sec:after-chaotic-tides}.

In the following sections, we elaborate on those pieces of physics and how they are integrated into our MBH-binary model.

\subsection{Outer orbit relaxation}\label{sec:Methods-outer-relaxation}
Consider a binary system on an eccentric outer orbit, with the specific angular momentum $\bs{L}$, energy $E_{\rm orb}$, and eccentricity vector $\bs{e}$.
Due to gravitational encounters with other stars, $\bs{L}$ and $E_{\rm orb}$ will gradually change. For highly eccentric orbits, the angular momentum relaxation timescale $t_{L,\text{relax}}$ is usually smaller than the energy relaxation timescale $t_{E,\text{relax}}$ \citep{Sari2019}. Furthermore, since the MBH perturbation on the inner orbit is determined mostly by $r_{p}$ that changes on the timescale $t_{L,\text{relax}}$, we only consider the relaxation of $\bs{L}$ and fix $E_{\rm orb}$ to its initial value.

In highly eccentric orbits, the position of the binary $\bs{r}$ is nearly parallel to the eccentricity vector $\bs{e}$ most of the time, so gravitational encounters causing velocity deflections mainly change $\bs{L}$ in the direction perpendicular to $\bs{e}$. More specifically, the angular momentum change $\Delta \bs{L}$ per outer orbit lies in the plane spanned by $\hat{\bs{L}}$ and $\hat{\bs{L}}\times\hat{\bs{e}}$, where $\hat{\bs{L}}$ and $\hat{\bs{e}}$ are the unit vector in the directions of $\bs{L}$ and $\bs{e}$. Given the period of the outer orbit $P(a)=2\pi\sqrt{a^3/G\MBH}$ and the two-body relaxation time $t_{\text{2B,relax}}$, we model $\Delta \bs{L}$ as follows\footnote{In \citet{Bradnick2017}, the diffusion of $\bs{L}$ is assumed to be isotropic. However, as argued above, for highly eccentric orbits it is more accurately modeled as being confined to the plane perpendicular to $\bs{e}$.}
\begin{equation}\label{eq:L_diffusion}
    \Delta\bs{L}=L_c(a)\sqrt{\frac{P(a)}{t_{\text{2B,relax}}}}\lrb{c_1\hat{\bs{L}}+c_2\hat{\bs{L}}\times\hat{\bs{e}}},
\end{equation}
where $L_c(a)=\sqrt{G\MBH a}$ is the angular momentum for a circular outer orbit, and $c_1,c_2$ are two random numbers drawn independently from the normal distribution with zero mean and unit variance.

The per-outer-orbit change in the magnitude of angular momentum $\Delta L$ can also be recovered from eq.~(\ref{eq:L_diffusion}):
\begin{align}
    \lara{\Delta L} = & \sum_i \frac{L_i}{L}\lara{\Delta L_i}\nonumber\\
    & +\sum_{ij}\frac{1}{2L}\lrb{\delta_{ij}-\frac{L_{i}L_{j}}{L^{2}}}\lara{\Delta L_{i}\Delta L_{j}}\nonumber\\
    = & \frac{L_c^2(a)}{L}\frac{P(a)}{t_{\text{2B,relax}}}, \\
    \lara{\lrb{\Delta L}^2} & = \sum_{ij}\frac{L_iL_j}{L^2}\lara{\Delta L_{i}\Delta L_{j}} = L_c^2(a)\frac{P(a)}{t_{\text{2B,relax}}}.
\end{align}
Therefore, the random walk of $\bs{L}$ as modeled by eq. (\ref{eq:L_diffusion}) automatically incorporates both the drift $\lara{\Delta L}$ and diffusion $\lara{\lrb{\Delta L}^2}$ of the magnitude of angular momentum $L$. 

In principle $t_{\text{2B,relax}}$ depends on the radial number density profiles of the stars and other compact objects at the Galactic Center, and is a function of radius $r$ \citep{Kocsis2011}. However, since those profiles are not well-constrained within the central parsec of the Galactic Center, we use a fixed value of $t_{\text{2B,relax}}$ for a given system (see Section~\ref{sec:Methods-summary}).
The required number of outer orbits to change $L$ significantly can be obtained from $t_{\text{2B,relax}}$:
\begin{align}\label{eq:N_l}
    N_l & \sim \frac{L^2}{\lara{\lrb{\Delta L}^2}} = \frac{L^2}{L_c^2(a)}\frac{t_{\text{2B,relax}}}{P(a)} \approx \frac{2r_p}{a}\frac{t_{\text{2B,relax}}}{P(a)} \nonumber\\
    & = 2.4\times10^3\lrb{\frac{a_b}{1\,\text{AU}}}\lrb{\frac{r_p}{70\,r_t}}\lrb{\frac{t_{\text{2B,relax}}}{1\,\text{Gyr}}}\lrb{\frac{a}{1\,\text{pc}}}^{-5/2}.
\end{align}
The corresponding angular momentum relaxation timescale is
\begin{align}\label{eq:t-relax-AM}
    t_{L,\text{relax}} = N_l P \approx \lrb{\frac{2r_p}{a}}t_{\text{2B,relax}}.
\end{align}

\subsection{MBH tidal perturbations}\label{sec:Methods-MBH-perturbation}
\begin{figure}
    \centering
    \includegraphics[width=\columnwidth]{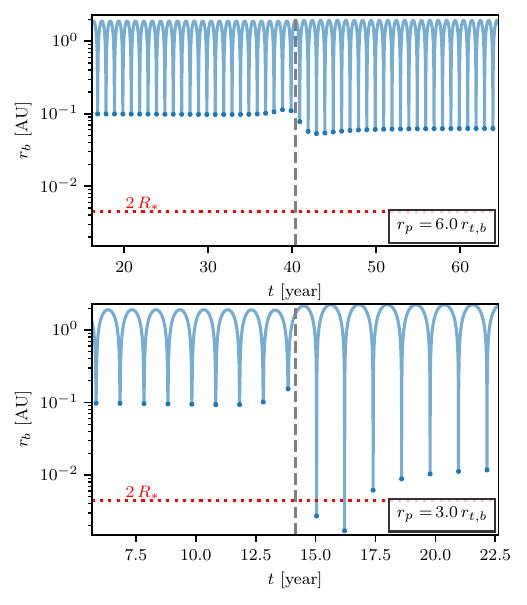}
    \vspace{-0.6cm}
    \caption{The binary separation $r_b(t)$ as a function of time. The binary system is on an eccentric orbit around the MBH, with $a=0.5\,\text{pc}$ and $r_p=6.0\,r_t$ (upper panel), $3.0\,r_t$ (lower panel). The inner orbit has initial SMA $a_{b,0}=1.0\,\text{AU}$ and eccentricity $e_{b,0}=0.9$. The outer orbit has $\bs{L}$ in $+\hat{z}$ direction and pericenter on $-\hat{x}$ axis. The initial orientation of the binary system are specified by the orbital elements $i_{b,0}=0.7\pi$ (inclination angle), $\omega_{b,0}=0.1\pi$ (argument of pericenter), $\Omega_{b,0}=0.95\pi$ (longitude of ascending node), and $M_{b,0}=1.0$ (mean anomaly). The gray vertical line marks the time of the outer pericenter passage, the red horizontal line is the sum of the two binary star's radii, and the inner pericenter radii are highlighted with circular markers. For $r_p=3.0\,r_t$ (bottom panel), the binary separation and inner pericenter radii drop below $2\,R_*$ temporarily near the outer pericenter, which highlights the sharp change in $e_b$ when $r_p$ is small. In our simulation, the binary system on the lower panel is flagged as collision as soon as the binary separation becomes less than $2\,R_*$.}
    \label{fig:binary_separation}
\end{figure}

During the outer pericenter passage, the binary's inner orbit is perturbed by the MBH tides. Depending on the orbit orientation, $e_b$ can either increase or decrease. The effect of the perturbation can be roughly divided into two regimes:
\begin{enumerate}
    \item For $r_p\gtrsim 10\,r_t$, the MBH perturbation primarily affects $e_b$ while keeping $a_b$ largely unchanged. Since the angular frequency of the inner orbit is much higher than that of the outer pericenter passage, the change in $e_b$ occurs gradually over many inner orbits, and the secular approximation can provide an analytical estimate of the total change in $e_b$ and in the inner orbit orientation \citep{Hamers2019}. We adapt the result of \citet{Hamers2019} in the parabolic encounter limit ($e=1$) to approximate our highly eccentric outer orbits. The full analytical expression of the inner orbit perturbation is given in Appendix~\ref{sec:App_SA_perturb}. If the outer orbit does not undergo relaxation, the repeated perturbations over multiple outer orbits become the eZLK effect, where $e_b$ can be driven to near unity \citep[see the review by][]{Naoz2016}. However, since the outer orbit of the binary system is constantly under relaxation, the change in $e_b$ is stochastic over many outer orbits and cannot be analytically described.
    \item For $r_p\lesssim 10\,r_t$, the effect of MBH tides on the binary system is increasingly chaotic. As the angular frequency of the outer pericenter passage becomes comparable to that of the inner orbit, the secular approximation breaks down. To resolve this complex three-body interaction and be conservative in our treatment, we numerically integrate the orbital dynamics during the outer pericenter passage whenever $r_p<30\,r_t$, using the N-body simulation code REBOUND\footnote{Our numerical integration does not include the relativistic effects because the Schwarzschild precessions operate on the timescale much larger than the outer pericenter passage. See Appendix~\ref{sec:App-precessions} for a discussion on the effects of Schwarzschild precessions.} with the IAS15 integrator \citep{Rein2012,Rein2015}. Since the tidal interaction between the MBH and the binary is strongly localized near the outer pericenter, we restrict the integration to the section of the orbit where the MBH-binary separation is less than $10\times\max(r_t,r_p)$. 
\end{enumerate}
Figure~\ref{fig:binary_separation} shows two examples of the evolution of binary separation $r_b(t)$ near one outer pericenter passage. Initially $r_b$ changes periodically due to the non-zero inner eccentricity. As the binary passes through the outer pericenter, the inner orbit gets perturbed by the MBH. For $r_p=6.0\,r_t$ (upper panel), the change in $e_b$ and $r_{p,b}$ is relatively smooth, while for $r_p=3.0\,r_t$ (lower panel), the change is more abrupt. Furthermore, for $r_p=3.0\,r_t$, the binary separation can briefly reach below $2\,R_{*}$ during the outer pericenter passage and lead to binary collision. This highlights the importance of resolving the three-body dynamics explicitly, since the collision can only be detected by constantly monitoring the binary throughout the MBH perturbation.
For $r_p \lesssim 3\,r_t$, the MBH perturbation can also result in the Hills breakup. This outcome is captured in our REBOUND simulations, where we measure the velocity $v_{\rm HVS}$ of the ejected star and the orbital properties of the bound star.

\subsection{Chaotic tides and inner orbit dynamics}\label{sec:Methods-chaotic-tides}

\begin{figure}
    \centering
    \includegraphics[width=\columnwidth]{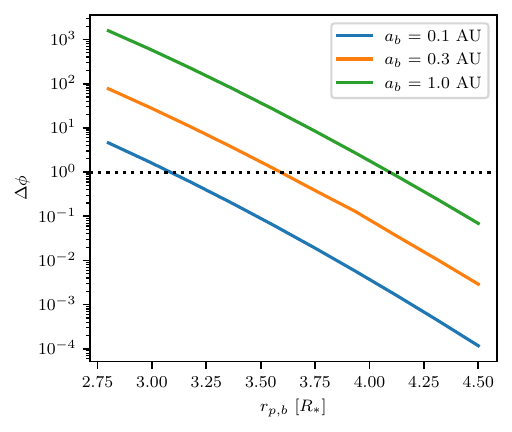}
    \vspace{-0.6cm}
    \caption{The change in the stellar oscillation phase (eq.~\ref{eq:chaotic-tides-phase-change}) due to the variation in inner orbital period after the energy kick from the inner pericenter passage. The horizontal line, $\Delta\phi=1$, is the crude requirement for triggering the chaotic tides. The intersections between this horizontal line and the curves are $r_{b,\rm ct}$. Depending on the value of $a_b$, $r_{b,\rm ct}$ ranges from 3 to 4\,$R_*$.}
    \label{fig:r_ct}
\end{figure}

The tidal interactions between binary stars can be significant when $r_{p,b}$ is small. One popular method of modeling the tidal interactions is to only consider equilibrium tides, where the stars are assumed to be in instantaneous hydrostatic equilibrium with the tidal field of the companions \citep{hut81_tidal_friction,Eggleton1998,Eggleton2001}.
Although this approach has been used in previous works \citep{Bradnick2017,stephan19_eLK}, it neglects the dynamical and oscillatory nature of tides.
Specifically, when $e_b$ is sufficiently high and $r_{p,b}$ is only a few times $R_*$, the tidal force of the companion can excite the stellar oscillation modes during the inner pericenter passage and transfer a fraction of the orbital energy to the oscillation modes. This energy transfer between orbit and stellar oscillation was first investigated in the tidal capture of binary systems \citep{Press1977,Kochanek1992,Mardling1995}. 
\citet{Mardling1995} demonstrated that when $r_{p,b}$ is sufficiently small, the modulation of the orbital period due to energy transfer will cause the oscillation mode to receive kicks at effectively random phases during pericenter passages. The random kicks in the stellar oscillation over multiple orbits will lead to its chaotic growth, during which the orbital energy is quickly drained and the binary is hardened efficiently. The efficient hardening from chaotic tides has been proposed to explain the migration of hot Jupiters \citep{Wu2018,Vick2018}. For high $e_b$, the tidal interaction is localized near the pericenter passage, and the mode energy evolution over multiple inner orbits can be described using an iterative map \citep{Vick2018}. In the following, we describe our use of the iterative map to solve the interactions between stellar oscillation modes and orbital dynamics self-consistently.

In our model we only consider a single stellar oscillation mode, the $l=m=2\,f$-mode, in Star 1, while ignoring all modes in Star 2 (to be conservative in our model). We use MESA \citep{Paxton2011, Paxton2013, Paxton2015, Paxton2018, Paxton2019, Jermyn2023} to produce the stellar model, and then post-process it with GYRE \citep{Townsend2013} to compute the eigenfunction and eigenfrequency of the stellar oscillation modes. The $l=m=2\,f$-mode has angular frequency $\omega_f$ and initial complex amplitude $A_f=0$. The amplitude is normalized so that the energy and angular momentum stored in the mode are $E_f=\abs{A_f}^2$ and $L_f=2\abs{A_f}^2/\omega_f$. Let the subscript $_k$ denotes the quantities right before the $k$-th inner pericenter passages. The iterative map of the complex amplitude $A_f$ is given by \citep{Vick2018}:
\begin{equation}\label{eq:iter-map}
    A_{f,k+1}=\lrb{A_{f,k}+\Delta A_{f,k}}\,e^{i\omega_f P_{b,k+1}},
\end{equation}
where $P_{b,k}$ is the period of the $k$-th inner orbit, and $\Delta A_k$ describes the kick from the tidal force during the $k$-th pericenter passage. The value of $\Delta A_{f,k}$ depends strongly on $r_{p,b}$ and weakly on $e_{b}$. A more detailed description of the iterative map and the expression for $\Delta A_{f,k}$ is provided in Appendix~\ref{sec:App-iter-map}.

\begin{table*}
	\centering
	\caption{Parameters adopted in our fiducial models.}
	\label{tab:model-parameters}
	\begin{tabular}{lcl}
		\hline
		Symbol & Values/Ranges & Description\\
		\hline
		$M_{*}$ & $0.5\,M_\odot$ & Mass of each star\\
		$R_{*}$ & $0.478\,R_\odot$ & Radius of each star\\
		$\bar{Q}$ & 0.1198 & Dimensionless tidal coupling coefficient for $l=m=2\,f$-mode\\
        $\omega_f$& $1.67084\sqrt{GM_{*}/R_{*}^3}$ & Angular frequency for for $l=m=2\,f$-mode\\
        $a_{b,0}$ & $0.1,0.3,1.0\,\text{AU}$  & Initial inner SMA\\
        $e_{b,0}$ & $\lrsb{0.1,1-8\,R_\odot/a_{b,0}}$ & Initial inner eccentricity\\
        $\eta$    & $0,1$ & Power law index of the inner eccentricity distribution\\
        $\MBH$    & $4.26\times10^6\,M_\odot$ & MBH mass\\
        $a$       & $0.5,2.0\,\text{pc}$ & Outer SMA\\
        $r_{p,0}$ & $70\,t_{t,b}$ & Initial outer pericenter radius\\
        $t_{\text{2B,relax}}$ & $1,10\,\text{Gyr}$ & Two-body relaxation time of the outer orbit\\
		\hline
	\end{tabular}
    \vspace{0.6cm}
\end{table*}

\begin{figure*}
    \centering
    \includegraphics[width=0.75\textwidth]{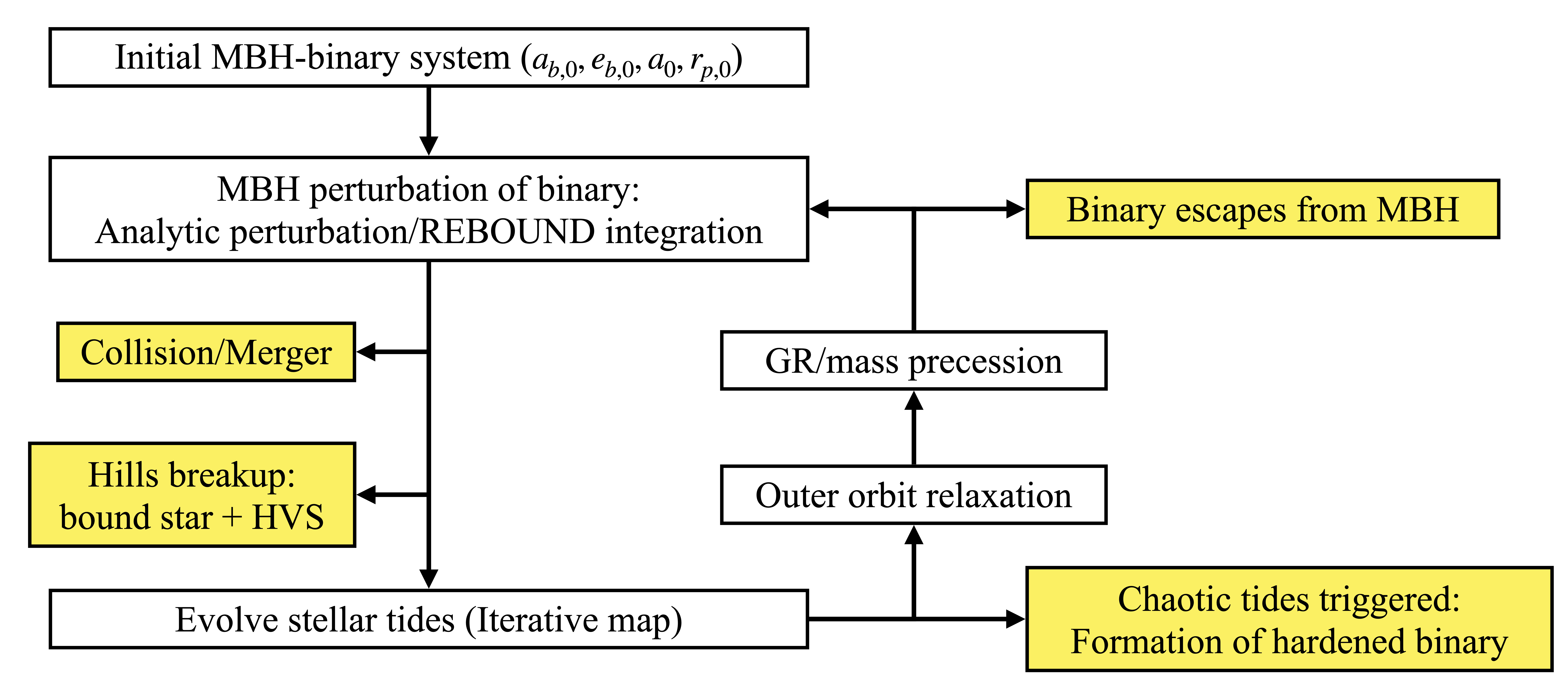}
    \caption{The simulation flow of our MBH-binary three-body system. The end states of the binary system are highlighted with yellow color. The simulation is stopped when the system reaches any of the end states, or when the total simulation time exceeds $0.1\,t_{\text{2B,relax}}$.}
    \label{fig:simulation_flow}
\end{figure*}

The chaotic growth in $A_{f}$ requires that consecutive kicks happen at random phases. With zero initial mode energy, this requirement is roughly given by \citep{Wu2018,Vick2018}:
\begin{equation}\label{eq:chaotic-tides-phase-change}
    1\mathrm{\,rad}<\Delta\phi = \omega_f \abs{P_{b,1}-P_{b,0}} = \frac{3}{2}\omega_fP_{b,0}\abs{\frac{\Delta E_{f,0}}{E_{b,\text{orb},0}}},
\end{equation}
where $E_{b,\text{orb},0}$ is the initial inner orbital energy and $\Delta E_{f,0}=\abs{\Delta A_{f,0}}^2$ is the initial energy kick. We assume that the fractional change in period is small ($|\Delta E_{f,0}|\ll |E_{b,\text{orb},0}|$) to obtain eq. (\ref{eq:chaotic-tides-phase-change}).
Figure~\ref{fig:r_ct} shows $\Delta\phi(r_{p,b})$ for $a_b=0.1,0.3,1.0\,\text{AU}$ for the first pericenter passage. The steepness of the function  reflects the sensitivity of $\Delta E_{f,0}$ to $r_{p,b}$. For a given $a_b$, we define the critical radius of chaotic tides $r_{b,\rm ct}$ as
\begin{equation}\label{eq:r-ct}
    \Delta\phi(r_{p,b}=r_{b,\rm ct})=1.
\end{equation}
The value of $r_{b,\rm ct}$ ranges from 3 to $4\,R_{*}$ for $0.1<a_b<1\,{\rm AU}$.
Note that $r_{b,\rm ct}$ does not represent a clear boundary of chaotic tides due to their stochastic nature \citep{Vick2018} and only serves as a characteristic scale. In practice, the chaotic regime may be reached for $r_{p,b}>r_{b,\rm ct}$, while some configurations within this radius may remain stable.

Once chaotic tides are triggered, $E_f$ will grow, on average, as
\begin{equation}
    \lara{E_f} \sim N_b\Delta E_{f,0},
\end{equation}
where $N_b$ is the number of inner orbits.
The large phase change $\Delta\phi$ required to trigger the chaotic tides (eq.~\ref{eq:chaotic-tides-phase-change}) can be translated into a minimum growth rate:
\begin{align}
    E_f \gtrsim N_b\cdot\frac{2}{3}\frac{\abs{E_{b,\text{orb},0}}}{\omega_fP_{b,0}}.
\end{align}
The number of inner orbits for $E_f$ to grow to $0.1\abs{E_{b,\text{orb},0}}$ is
\begin{align}
    N_b & \lesssim 0.15\,\omega_fP_{b,0} \nonumber\\
    & = 1.1\cdot10^4\lrb{\frac{a_b}{1.0\,\text{AU}}}^{3/2}.
\end{align}
As we mainly consider binaries on wide outer orbits with $P/P_b=4.5\times 10^4(a/{\rm pc})^{3/2}(a_b/{\rm AU})^{-3/2}$, we find that chaotic tides, once triggered, will most likely shrink the inner orbit within one outer orbit. If so, the inner orbit evolution will rapidly decouple from the outer orbit --- this rapid decoupling is not correctly captured in previous works \citep{Bradnick2017, stephan19_eLK}.

In our model, we check for the onset of chaotic tides once per outer orbit using the iterative map. After applying the MBH perturbation at the outer pericenter (Section~\ref{sec:Methods-MBH-perturbation}), the updated $a_b$ and $e_b$ serve as initial conditions for the map (eq.~\ref{eq:iter-map}) to evolve $E_f$. The maximum number of iteration steps is (conservatively) set to $0.9\times P/P_b$ inner orbits. When $E_f$ exceeds $0.3\abs{E_{b,\text{orb},0}}$, we consider the binary to have undergone significant orbital shrinkage and triggered chaotic tides. Our results are insensitive to the exact choice of $0.3\abs{E_{b,\text{orb},0}}$, because the mode energy grows extremely rapidly when chaotic tides are triggered. The subsequent orbital evolution is not included in our model, but is discussed in Sections~\ref{sec:chaotic-tides-damping} and~\ref{sec:after-chaotic-tides}.

\subsection{Model parameters and potential outcomes}\label{sec:Methods-summary}
The parameters used in our fiducial models are listed in Table~\ref{tab:model-parameters}.
Our MBH-binary model incorporates the physics of two-body relaxation, tidal interaction between two stars, and MBH perturbation of the inner orbit. We also include the effects of Schwarzschild and mass precessions, which are described in detail in Appendix~\ref{sec:App-precessions}. We consider three choices of initial inner SMA $a_{b,0}=0.1,0.3,1.0\,\text{AU}$. For each $a_{b,0}$, we further select initial inner eccentricity $e_{b,0}$ from nine evenly spaced values in the range $\lrsb{0.1,1-8\,R_\odot/a_{b,0}}$. This avoids very small $r_{p,b}<8\,R_\odot$ and prevents our statistics from being biased by those binaries that are already close to collision/triggering of chaotic tides.
The initial inner orbital orientation and phase are determined by the longitude of ascending node $\Omega_b$, inclination $i_b$, the argument of pericenter $\omega_b$, and mean anomaly $M_b$, for which we randomly sample from uniform distributions in the ranges $\Omega_b\in[0,2\pi],\cos i\in[-1,1],\omega_b\in [0,2\pi],M_b\in [0,2\pi]$.
The outer orbit is initialized in the $xy$ plane, with the angular momentum in $+\hat{z}$ direction and pericenter on the $-\hat{x}$ axis.

\begin{figure*}
    \centering
    \includegraphics[width=\textwidth]{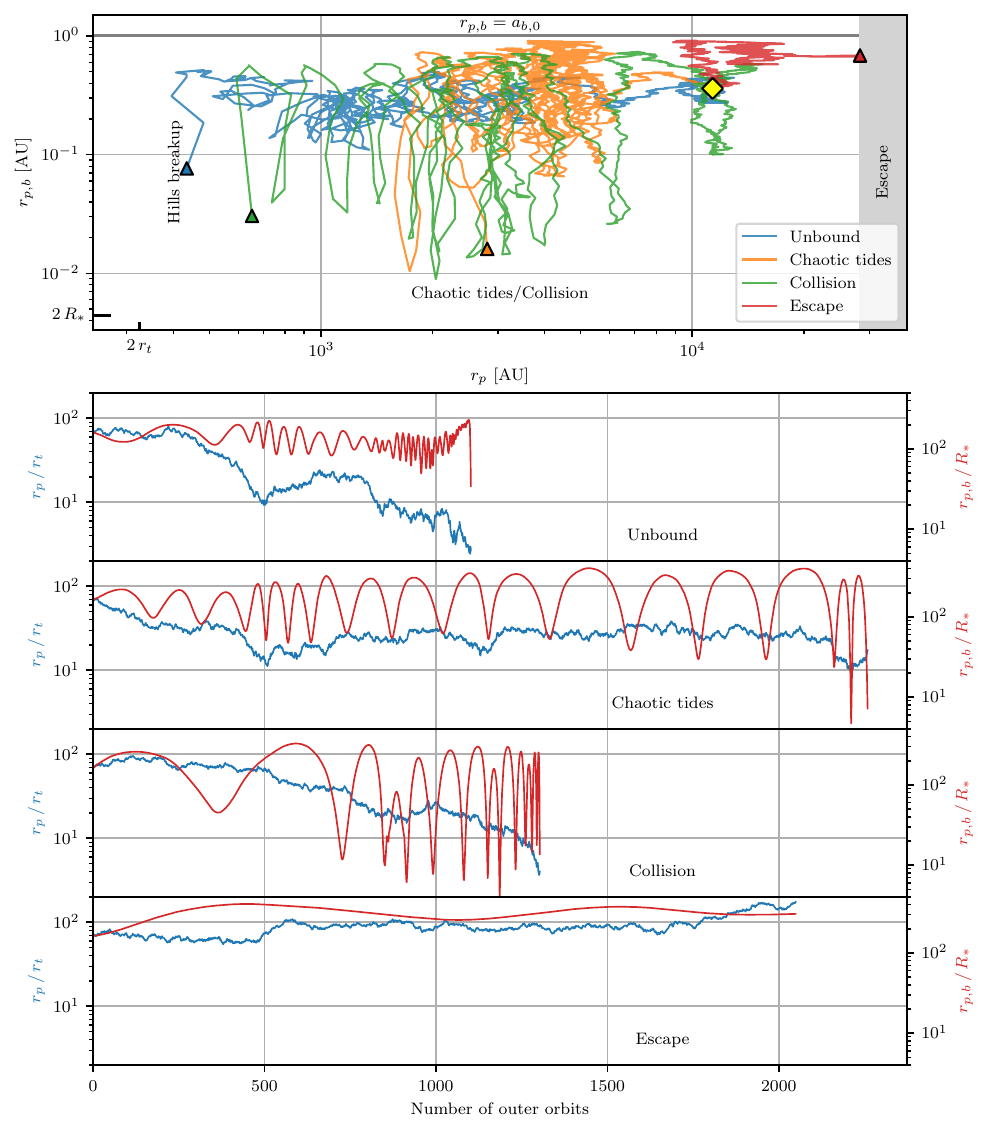}
    \vspace{-0.6cm}
    \caption{Example trajectories of the MBH-binary system. All systems have the same initial conditions $(a,t_{\text{2B,relax}},a_{b,0},e_{b,0})=(0.5\,\text{pc},1\,\text{Gyr},1.0\,\text{AU},0.64)$. \textit{Top panel}: The evolution of $r_p$ and $r_{p,b}$. The initial state of the systems is marked with the yellow diamond. The states of the systems prior to the final outer pericenter passage are highlighted with triangles. \textit{Bottom four panels}: $r_p$ and $r_{p,b}$ evolution as the function of outer orbit number for the same trajectories as in the top panel. The random walk in $r_p$ and the eZLK-like oscillation can be clearly seen.}
    \label{fig:binary-trajectories}
\end{figure*}

We consider two choices of outer SMA $a=0.5,2.0\,\text{pc}$. The initial outer pericenter radius $r_{p,0}$ is fixed to $70\,r_t$ to ensure that the MBH's influence on the inner orbit is initially negligible. Two relaxation timescales $t_{\text{2B,relax}}=1,10\,\text{Gyr}$ are used to account for the uncertainties in the density profiles at the Galactic Center. 

For each set of $(a,t_{\text{2B,relax}},a_{b,0},e_{b,0})$, 1000 binary samples are simulated to collect the statistics of different outcomes. Figure~\ref{fig:simulation_flow} illustrates the flow of the simulation cycle. The simulation steps are summarized as follows:
\begin{enumerate}
    \item Initialize the inner orbit and outer orbit.
    \item Simulate each outer pericenter passage and the MBH perturbation on the inner orbit. Depending on the size of $r_p$, this is handled differently:
    \begin{itemize}
        \item For $r_p>30\,r_t$, the pericenter passage is not resolved and the perturbation to the inner orbit is computed with analytical secular approximation (see Section~\ref{sec:Methods-MBH-perturbation} and Appendix~\ref{sec:App_SA_perturb}).
        \item For $r_p<30\,r_t$, the three-body interaction during the passage is numerically integrated with REBOUND package \citep{Rein2012,Rein2015}. This simulation is stopped if the binary stars collide or undergo Hills breakup.
    \end{itemize}
    \item Use the iterative map to evolve the stellar oscillation mode (see Section~\ref{sec:Methods-chaotic-tides}). The simulation is stopped if the chaotic tides are considered to be triggered.
    \item Relax the outer orbit according to the method in Section~\ref{sec:Methods-outer-relaxation}. If $r_p>175\,r_t$ after the relaxation, the binary system is considered to be too far from the MBH and the simulation stops.
    \item Update the inner and outer arguments of pericenter $\omega_b,\omega$ with the Schwarzschild and mass precession, respectively (see Appendix~\ref{sec:App-precessions}).
    \item Repeat Step (2) to (5) until one of the stopping conditions is reached, or when the total simulation time reaches $0.1\,t_{\text{2B,relax}}$
\end{enumerate}

\section{Results}\label{sec:Results}

In this section we present our results from Monte Carlo simulations. We first focus on the evolution of individual systems over multiple outer orbits, and illustrate the working of chaotic tides. We then show the statistics of the different outcomes, and their dependence on the initial orbital properties and the relaxation time. Finally, we present the statistics on the stellar collisions and the products of Hills breakup, including the orbital properties of HVS and bound stars.

\subsection{Trajectories of MBH-binary system}

We start by presenting the simulation results for individual systems. Figure~\ref{fig:binary-trajectories} shows four example ($r_p,r_{p,b}$) trajectories. All systems have $(a,t_{\text{2B,relax}},a_{b,0},e_{b,0})=(0.5\,\text{pc},1\,\text{Gyr},1.0\,\text{AU},0.64)$. The four trajectories correspond to the four possible end states of the system, "Unbound from Hills breakup," "Trigger of chaotic tides," "Collision," and "Escape from MBH" (see also Figure~\ref{fig:simulation_flow}). Qualitatively, the trajectory can be described as the combination of the random walk in $r_p$ due to the outer orbit relaxation, and the the eZLK-like movement in $r_{p,b}$ due to the MBH perturbation. If the former prevails, the binary will either undergo Hills breakup ($r_p\lesssim3\,r_t$) or escape from the tidal influence of the MBH ($r_p>175\,r_t$). Contrarily, if the latter excites $e_b$ to a large value, the binary system may collide or experience chaotic tides. The end state of the binary system is controlled by the competition between these two stochastic processes. For most part of the trajectories, the change in $a_b$ is negligible. However, significant change in $a_b$ is still possible when $r_p\lesssim\text{a few}\times r_t$ \citep{Zhang2010}.

\subsection{Chaotic tides evolution}

\begin{figure*}
    \centering
    \includegraphics[width=\textwidth]{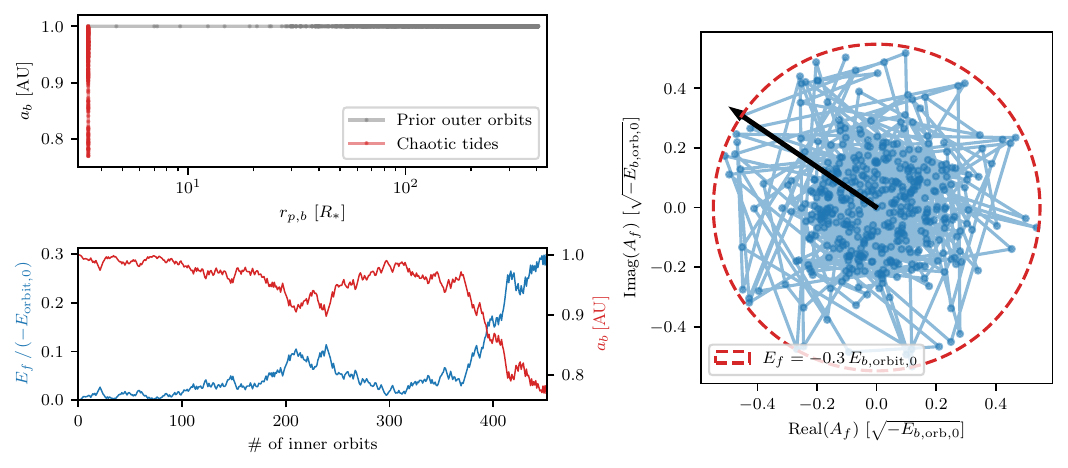}
    \vspace{-0.6cm}
    \caption{A binary system that undergoes chaotic tides. The system has initial conditions $(a,t_{\text{2B,relax}},a_{b,0},e_{b,0})=(0.5\,\text{pc},1\,\text{Gyr},1.0\,\text{AU},0.64)$. \textit{Upper left panel:} The evolution of $a_b$ and $r_{p,b}$ over the multiple outer orbits (gray line) and over the final outer orbit when chaotic tides are triggered (red line). The dots indicate the values before each outer/inner pericenter passage. \textit{Lower left panel:} The evolution of $E_f$ (blue line, left axis) and $a_b$ (red line, right axis) over multiple inner orbits after the final outer pericenter passage. \textit{Right panel:} The evolution of the complex amplitude $A_f$ during the growth of chaotic tides. The axis scales are normalized such that $|A_f|=1$ corresponds to the initial inner orbital binding energy $|E_{b,\text{orb,0}}|$. The red dashed circle marks our criterion for the trigger of the chaotic tides, $E_f=0.3\times |E_{b,\text{orb,0}}|$. The black arrow highlights the state of the stellar oscillation at the end of our simulation.}
    \label{fig:chaotic-tides-example}
\end{figure*}

Figure~\ref{fig:chaotic-tides-example} shows the evolution of a binary system that undergoes chaotic tides. The upper left panel plots $a_b$ versus $r_{p,b}$ over multiple outer orbits and during the growth of chaotic tides. Before the final outer pericenter passage, the MBH repeatedly perturbs $e_b$ with little change in $a_b$. This is reflected in the horizontal gray line. During the final outer pericenter passage, the MBH perturbs $r_{p,b}$ to $\sim3.5\,R_*$ and triggers the chaotic growth of stellar tides. The rapid growth of stellar tides shrinks $a_b$ while keeping $r_{p,b}$ mostly unchanged, as shown by the red vertical line. The evolution of $a_b$ and $E_f$ during the chaotic growth of tides is shown in the lower left panel. It is clear that the increase in $E_f$ is accompanied by the decrease in $a_b$ due to the conservation of the total energy, and that the growth in $E_f$ is not monotonic but diffusive. The right panel offers a more detailed look of the iterative map through the evolution of $A_f$ (eq.~\ref{eq:iter-map}).

\subsection{Statistics of different binary outcomes}\label{sec:outcome-statistics}
We next discuss the statistics of the three possible end states (excluding ``Escaped''; Figure~\ref{fig:simulation_flow}) and their dependence on the initial conditions. We show that the emptiness of the loss cone provides a good indicator of the outcome statistics. We also examine the final $r_p$ and demonstrate that chaotic tides can be triggered for $r_p\gtrsim\text{a few}\times r_t$, where the MBH perturbation is still weak.

\subsubsection{Empty versus full loss cone}
\begin{figure}
    \centering
    \includegraphics[width=\columnwidth]{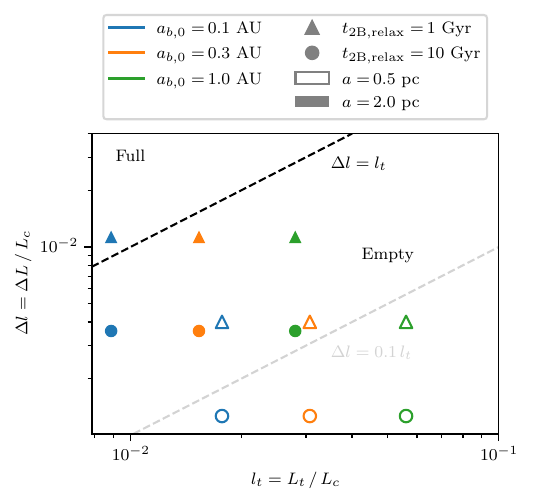}
    \vspace{-0.6cm}
    \caption{The per-outer-orbit angular momentum dispersion $\Delta l$ versus the angular momentum at the binary disruption radius $l_t$. Both $\Delta l$ and $l_t$ (eq.~\ref{eq:delta_l_ltb}) are dimensionless and normalized with $L_c=\sqrt{G\MBH a}$. The system properties $(a_{b,0},t_{\text{2B,relax}},a)$ are indicated with different markers. The black dashed line highlights $\Delta l=l_t$, the rough boundary between empty and full loss cone regime.
    }
    \label{fig:delta_l_l_tb}
\end{figure}

The efficiency of angular momentum relaxation determines the number of outer orbits before the binary system reaches $r_p\sim r_t$ or escapes the influence of the MBH tides. This can be quantified by $\Delta l\equiv \Delta L/L_c$, the per-outer-orbit dispersion in the dimensionless angular momentum, and $l_t$, the dimensionless angular momentum at the binary tidal breakup radius:
\begin{align}\label{eq:delta_l_ltb}
    \Delta l & \equiv \Delta L/L_c = \sqrt{\frac{P}{t_{\text{2B,relax}}}},\\
    l_t & \equiv L_t/L_c = \sqrt{1-\lrb{1-\frac{r_t}{a}}^2},
\end{align}
where $L_c = \sqrt{GMa}$ is the angular momentum of a circular outer orbit, $\Delta L$ is the per-outer-orbit dispersion in the outer orbital angular momentum due to relaxation, and $L_t= \sqrt{GMa [1 - (1-r_t/a)^2]}\approx \sqrt{2GMr_t}$ is the critical angular momentum at which the binary will be tidally disrupted.
For $\Delta l> l_t$, the binary system is in the full loss cone regime and $r_p$ approaches $r_t$ abruptly. For $\Delta l< l_t$, the binary system is in the empty loss cone regime and $r_p$ approaches $r_t$ in a gradual and diffusive way. Figure~\ref{fig:delta_l_l_tb} shows $\Delta l$ versus $l_t$ for systems of different parameters (see Table~\ref{tab:model-parameters}). The black line indicates the boundary between empty and full loss cone regimes. Most of the systems in our model are in the empty loss cone regime, with a few of them close to the full loss cone regime.

\subsubsection{Outcome fractions}
\begin{figure}
    \centering
    \includegraphics[width=\columnwidth]{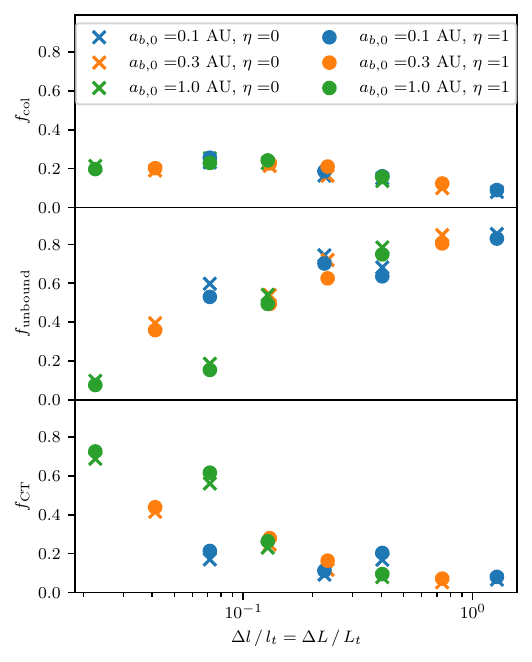}
    \vspace{-0.6cm}
    \caption{The fractions of different outcomes. The top, middle, and bottom panels show the fractions for collision $f_{\text{col}}$, unbound $f_{\text{unbound}}$ (Hills breakup), and chaotic tides $f_{\text{CT}}$, respectively. The horizontal axis, $\Delta l/l_t$ measures the emptiness of the loss cone. The statistical uncertainties from Monte Carlo simulations are estimated to be less than 4\% ($2\sigma$) and are not shown.
    \vspace{0.6cm}
    }
    \label{fig:outcome-fractions}
\end{figure}

We now present the statistics of different binary system outcomes. For a given set of $(a,t_{\text{2B,relax}},a_{b,0})$, we weigh the cases from different initial inner eccentricities $e_{b,0}$ according to a power-law distribution:
\begin{equation}\label{eq:eb-distribution}
    p(e_{b,0})\propto e_{b,0}^\eta,
\end{equation}
where $\eta$ is the power-law index. Two choices of $\eta$ are used: 0 for a uniform distribution and 1 for the thermal distribution, although our results depend weakly on the choice of $\eta$ (see later). Our statistics exclude the systems that are flagged as ``escaped'' or do not reach any end states after $0.1\,t_{\text{2B,relax}}$, which comprise $64\sim87$ percents of the sample.  Figure~\ref{fig:outcome-fractions} shows the fractions of three different outcomes --- collision $f_{\text{col}}$, unbound $f_{\text{unbound}}$ (Hills breakup), and chaotic tides $f_{\text{CT}}$. The fractions are normalized such that $f_{\text{col}}+f_{\text{unbound}}+f_{\text{CT}}=1$. The outcome fractions are presented as a function of $\Delta l/l_t$, which quantifies the emptiness of the loss cone. Several features can be observed from Figure~\ref{fig:outcome-fractions}:
\begin{enumerate}
    \item Independent of $\Delta l/l_t$ and $a_{b,0}$, stellar collisions are always subdominant, with $f_{\text{col}}$ ranging from 10\% to 25\%. A similar fraction was found in \citet{Sersante2025}.
    \item For $\Delta l/l_t$ close to unity (full loss-cone regime), the Hills breakup dominates the outcome for all $a_{b,0}$ ($f_{\text{unbound}}\gtrsim70\%$). 
    \item With decreasing $\Delta l/l_t$ (emptier loss-cone), there is a significant fraction of chaotic tides for $a_{b,0}\gtrsim 0.3\,\rm AU$. For $a_{b,0}=1.0\,\text{AU}$, $f_{\text{CT}}$ can reach above 60\% for $\Delta l/l_t<0.1$. For small $a_{b,0}$ (very close binaries), $f_{\text{CT}}$ remains low even for $\Delta l/l_t$ much less than unity.
    \item The outcome fractions are insensitive to the power index $\eta$ of the initial inner eccentricity distribution (eq.~\ref{eq:eb-distribution}), with little difference between $\eta=1$ (thermal) and 0 (uniform).
\end{enumerate}
The trends of $f_{\text{unbound}}$ and $f_{\text{CT}}$ with $\Delta l/l_t$ can be explained as follows. When $\Delta l/l_t$ is close to unity, $r_p$ approaches $r_t$ in only a few outer orbits and the cumulative MBH perturbation of $e_b$ is weak. Therefore, most systems that do not escape will reach $r_p\sim \text{a few}\times r_t$ and undergo Hills breakup.
On the other hand, when $\Delta l/l_t$ is much less than unity, the binary system takes a large number of outer orbits before $r_p$ can diffuse to the small scale of $r_t$. This leads to a large cumulative MBH perturbation on $e_b$ and increases the probability of the triggering of chaotic tides (and to a lesser extent the probability of binary collisions).
Since the triggering of chaotic tides requires $r_{p,b}\lesssim r_{b,\rm ct}$, $f_{\text{CT}}$ is the highest for $a_{b,0}=1.0\,\rm AU$, which has the largest $r_{b,\rm ct}$ and the least stringent requirement for chaotic tides (see Section~\ref{sec:Methods-chaotic-tides}). For $a_{b,0}=0.1\,\rm AU$, $f_{\text{CT}}$ remains low since the triggering of chaotic tides requires the binary to be near contact at the pericenter.

\subsubsection{Final pericenter radius}
\begin{figure}
    \centering
    \includegraphics[width=\columnwidth]{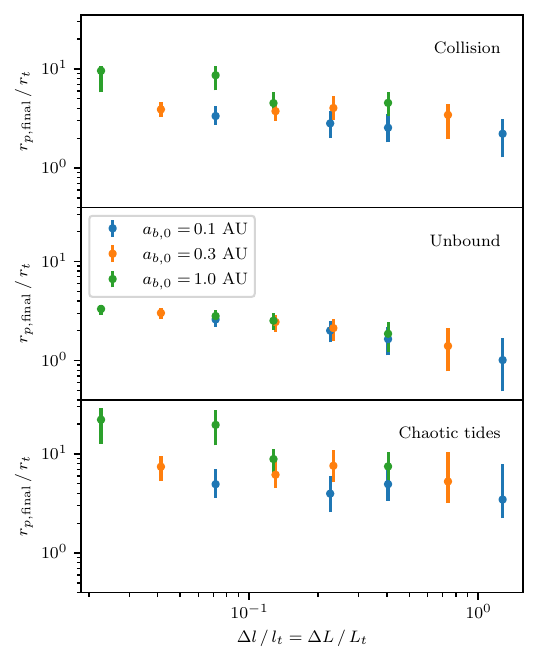}
    \vspace{-0.6cm}
    \caption{The final outer pericenter radius $r_{p,\text{final}}$ for different outcomes. The top, middle, and bottom panels show the cases of collision, unbound (Hills breakup), and chaotic tides, respectively. The horizontal axis, $\Delta l/l_t$, measures the emptiness of the loss cone. For each data point, the center dot is the median value in the simulation, and the error bar shows the extent from the first to third quartiles. Each simulated system is weighted assuming an initial thermal binary population, with $p({e_{b,0}})\propto e_{b,0}^\eta$ and $\eta=1$.}
    \label{fig:rp-final}
\end{figure}

The final outer pericenter radius $r_{p,\text{final}}$ before the binaries reach their end states reflects the nature of the outcomes.
Figure~\ref{fig:rp-final} shows $r_{p,\text{final}}$ for different binary outcomes as a function of $\Delta l/l_t$ (similar to Figure~\ref{fig:outcome-fractions}), assuming a thermal distribution $\eta=1$. Overall, the Hills breakup happens at the smallest radii $r_{p,\text{final}}\lesssim 3\,r_t$, which can be explained by the fact that unbinding the inner orbit requires strong MBH tidal forces that are only achieved for small $r_p$. Furthermore, $r_{p,\text{final}}$ is smaller in the full loss cone regime ($\Delta l\sim l_t$) than in the full loss cone regime ($\Delta l\ll l_t$). This is because faster relaxation and larger $\Delta l/l_t$ enable the binaries to penetrate deeper into the loss cone. For the chaotic tides outcome, our simulations show that they occur at larger radii than Hills breakup, with $r_{p,\text{final}}\sim 10\,r_t$. The larger values of $r_{p,\text{final}}$ for the chaotic tides are because triggering them requires a more gradual MBH perturbation to avoid binary destruction through collision or Hills breakup. Note that $r_{p,\text{final}}$ for $\Delta l/l_b\sim1$ is not reliable for chaotic tides due to very few such systems in our simulated sample.  For binary collisions, $r_{p,\text{final}}\sim3-10\,r_t$ is in between those of Hills breakup and chaotic tides.

\subsection{Products of Hills breakup}
In this subsection we present the statistics of the bound stars and HVS from our REBOUND simulations of the Hills breakup.
\subsubsection{Hyper-velocity stars}
\begin{figure}
    \centering
    \includegraphics[width=\columnwidth]{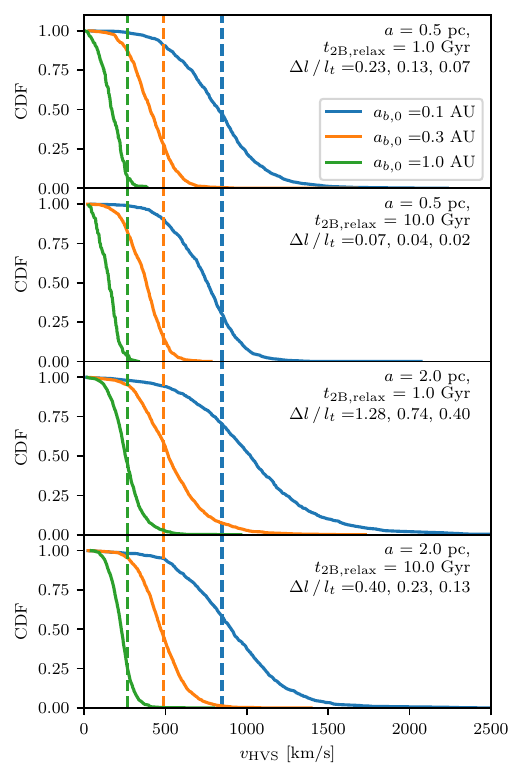}
    \vspace{-0.6cm}
    \caption{The cumulative distribution functions (CDFs) of $v_{\text{HVS}}$ from the Hills breakup. Each panel shows the result for a pair of $a$ and $t_{\text{2B,relax}}$, labeled in the top right corner. The listed values of $\Delta l / l_t$ correspond, from left to right, to $a_{b,0}=0.1,,0.3,$ and $1.0\,{\rm AU}$. The CDFs are constructed by weighting each simulated system with the thermal distribution of $e_{b,0}$ ($\eta=1$). Note that $v_{\text{HVS}}$ only considers the MBH potential and does not include the effects of the galactic potential. The vertical dash lines mark the analytical estimate given by eq.~(\ref{eq:v-HVS}).}
    \label{fig:v-HVS}
\end{figure}

Ignoring the galactic potential, the HVS ejected by the MBH through the Hills breakup have terminal velocities $v_{\rm HVS}$ on the order of
\begin{align}\label{eq:v-HVS}
    v_{\text{HVS}} & \sim \sqrt{\frac{GM_b}{2a_b}}\left(\frac{M_{\mathrm{BH}}}{M_b}\right)^{1/6}.
\end{align}
Figure~\ref{fig:v-HVS} shows the cumulative distribution functions (CDFs) of $v_{\text{HVS}}$ for different $a_{b,0}$. The rare cases where the ejected stars remain bound to the MBH are excluded. Each panel displays a combination of $a$ and $t_{\text{2B,relax}}$, and the statistics are weighted by the thermal distribution of $e_{b,0}$ ($\eta=1$). The values of $v_{\text{HVS}}$ from our REBOUND integrations agree with the analytical estimate (vertical dashed lines; eq.~\ref{eq:v-HVS}), although with a large spread due to variations in the system orientations and $r_{p,\rm final}$. The size of the spread in $v_{\text{HVS}}$ is related to the emptiness of the loss cone. For the empty loss cone cases, all the Hills breakups happen at similar outer pericenter radii, which results in smaller spread in $v_{\text{HVS}}$. For the full loss cone cases, the binary can reach radii much smaller than $3\,r_t$, where the stronger MBH tidal forces allow some stars to reach higher $v_{\text{HVS}}$. The second and third panels of Figure~\ref{fig:v-HVS} represent the empty and (nearly) full loss cones, respectively (see Section~\ref{sec:outcome-statistics}). The CDFs for the full loss cone have longer tails at velocities above the analytical estimate (eq.~\ref{eq:v-HVS}).

One prominent example of HVS of Galactic Center origin is S5-HVS1, which is believed to be a $\approx2.35\,M_\odot$ main-sequence star ejected through the Hills breakup at speed of $v_{\rm ej}\approx 1800\,\text{km}\,\text{s}^{-1}$ \citep{Koposov2020}. If we assume an equal-mass binary with $M_b=4.7\,M_\odot$ and $a_b=0.3\,\text{AU}$, this ejection speed is about 2.2 times the analytical prediction of $820\,\text{km}\,\text{s}^{-1}$ (eq.~\ref{eq:v-HVS}). Given our simulation results (Figure~\ref{fig:v-HVS}), a HVS with $v_{\text{HVS}}$ a factor 2 greater than the analytic estimate is rare, and we therefore infer that S5-HVS1 originated from a binary system with $a_b<0.3\,\text{AU}$ --- such a close binary may be primordial or the outcome of chaotic tides in the past.

\subsubsection{Stars bound to the MBH}
\begin{figure}
    \centering
    \includegraphics[width=\columnwidth]{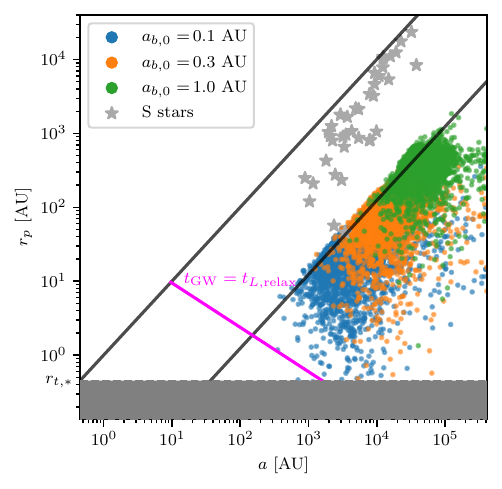}
    \vspace{-0.6cm}
    \caption{The orbital properties of the bound stars after the Hills breakup. The black lines show the constant eccentricity $e=0,1-2(M_b/\MBH)^{1/3}$. The bottom shaded region is where a single star will get tidally disrupted. The orbits of known S-stars are shown in gray star markers. The magenta line marks rough boundary between the dominance of GW emission/relaxation, assuming $t_{\text{2B,relax}}=1\,$Gyr (see eq.~\ref{eq:GW-relaxation-boundary}). Note that in this figure the data from different $a,t_{\text{2B,relax}}$ are combined, and that the distribution of bound stars should not be inspected too closely.
    \vspace{0.6cm}
    }
    \label{fig:bound-star}
\end{figure}

The other star from the Hills breakup remain bound to the MBH on a highly eccentric orbit:
\begin{equation}
    1-e\lesssim \lrb{\frac{M_b}{\MBH}}^{1/3}.
\end{equation}
Figure~\ref{fig:bound-star} shows the orbits of the bound stars from our simulations. The two black lines correspond to $e=0,1-2(M_b/\MBH)^{1/3}$. Most of the bound stars in our simulations have $e>1-2(M_b/\MBH)^{1/3}=0.988$. One constraint on the bound star orbits is the stellar tidal disruption radius:
\begin{equation}
    r_{t,*} = R_*\lrb{\frac{\MBH}{M_*}}^{1/3}.
\end{equation}
Below $r_{t,*}$, the star is destroyed in a TDE, although the exact radius at which this happens depends on the stellar density profile \citep[see e.g.,][]{Guillochon2013,Ryu2020}. This inaccessible region of the orbital properties for the bound stars is marked in gray.
In our simulations, a small fraction of binary systems that undergo Hills breakup in the full loss cone regime are deep plunging ($r_p<r_{t,*}$), and are expected to directly undergo double TDEs \citep{mandel15_dTDEs, yu24_tidal_break-up}. These systems do not form bound orbits around the MBH and are excluded from Figure~\ref{fig:bound-star}.

In Figure~\ref{fig:bound-star} the known S-stars at the Galactic Center are plotted in gray star markers \citep{Gillessen2017}. The low eccentricity of S-stars necessitates additional relaxation processes, if they originate from the Hills breakup \citep{Perets2009,generozov20_S_stars}. There are two main mechanisms to change the bound stars orbit --- gravitational wave (GW) orbital decay and the gravitational relaxation. The GW orbital decay shrinks the SMA on the timescale \citep{Peters1964,Sari2019}
\begin{equation}\label{eq:GW-timescale}
    t_{\text{GW}} \sim \frac{r_s}{c}\frac{\MBH}{M_*}\lrb{\frac{r_p}{r_s}}^4\lrb{\frac{a}{r_p}}^{1/2},
\end{equation}
where $r_s=2G\MBH/c^2$ is the Schwarzschild radius. The GW emission and gravitational relaxation modify the highly eccentric orbits in different ways --- while GW emission shrinks the SMA and keeps the pericenter radius roughly fixed, the relaxation mainly perturbs the angular momentum and the pericenter radius (see Section~\ref{sec:Methods-outer-relaxation}). With only GW orbital decay, the bound stars can be circularized and form the extreme mass ratio inspiral (EMRI). On the contrary, with only the relaxation, the bound stars cannot be circularized and may become TDEs or S-stars \citep{Sari2019,Perets2009}. The boundary of $t_{\text{GW}}=t_{L,\rm relax}$ separates the regions in which these two processes dominates:
\begin{align}\label{eq:GW-relaxation-boundary}
    \frac{r_{p}}{r_{s}} \sim &\, \lrb{2\,t_{\text{2B,relax}}\frac{M_{*}}{M_{\text{BH}}}\frac{c}{r_{s}}}^{2/5}\lrb{\frac{a}{r_{s}}}^{-3/5}\nonumber\\
    \sim &\, 2.0\times10^{3}\lrb{\frac{a}{r_{s}}}^{-3/5}\lrb{\frac{t_{\text{2B,relax}}}{10^{9}\,\text{yr}}}^{2/5}\nonumber\\
    &\times\lrb{\frac{M_{*}}{0.5\,M_{\odot}}}^{2/5}\lrb{\frac{M_{\text{BH}}}{4.26\times10^{6}\,M_{\odot}}}^{-7/5}.
\end{align}
This boundary is shown in Figure~\ref{fig:bound-star} as a magenta line, with $t_{\text{2B,relax}}=1\,$Gyr.\footnote{In reality, $t_{\text{2B,relax}}$ may weakly depends on the radius $r$ ($t_{\text{2B,relax}}\propto r^{-1/4}$ for a Bahcall-Wolf cusp \citep{Bahcall1976}). This dependence is ignored in our approximation, which makes our boundary slightly different from \citet{Sari2019}.}
As most of the bound orbits in our simulation are above this boundary and have $t_{\text{GW}}>t_{L,\rm relax}$, their orbits are mainly affected by relaxation (which changes $r_p$) and are unlikely to become EMRI. The production of EMRI requires the Hills breakup of tighter binaries with $a_b<0.1\,\rm AU$.

We emphasize that while our analysis above focuses on the relaxation from two-body encounters, there may be other more efficient mechanisms that can relax the orbits of the bound stars on shorter timescales \citep[e.g., scalar resonant relaxation; see][]{generozov20_S_stars}. It is more appropriate to think of our relaxation time $t_{\rm 2B,relax}$ as a timescale that encompasses all the uncertain relaxation mechanisms.

\subsection{Stellar collisions}
Stellar collisions comprise a small but not negligible fraction of simulation outcomes, regardless of the emptiness of the loss cone (see Figure~\ref{fig:outcome-fractions}).
Since systems entering the region $2\,R_*<r_{p,b}\lesssim r_{b,\rm ct}$ generally trigger chaotic tides, most stellar collisions require a strong kick from the MBH perturbation at $r_p\lesssim 10\,r_t$ to jump across the chaotic tide regime.
In addition to the collisions expected from the double-averaged orbital perturbation theory \citep{Hamers2019}, the complex three-body interactions between the MBH and binary can also occasionally drive $r_{p,b}$ below $2\,R_*$ transiently during the outer pericenter passage, even in cases where perturbation theory would suggest otherwise (see Figure~\ref{fig:binary_separation}).

\begin{figure*}
    \centering
    \includegraphics[width=\textwidth]{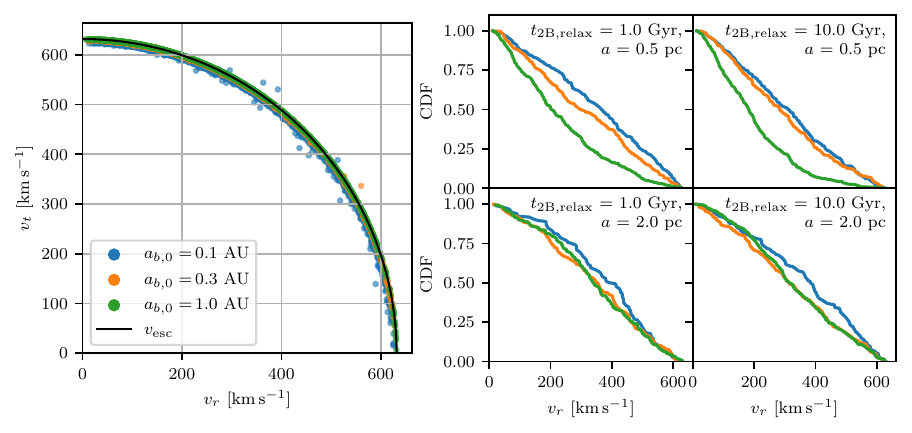}
    \vspace{-0.6cm}
    \caption{The collision velocity of the binaries. \textit{Left panel}: The tangential versus radial collision velocity ($v_t$ vs. $v_r$). The escape velocity $v_{\text{esc}}$ is highlighted with the black line. \textit{Right panel}: The CDF of $v_r$ assuming the thermal distribution of $e_b$ ($\eta=1$). The four cases correspond to different $t_{\text{2B,relax}}$ and $a$.}
    \label{fig:v-collide}
\end{figure*}

In our REBOUND integration, both the radial and tangential velocities, $v_r$ and $v_t$, at the moment of stellar collision are tracked. The left panel of Figure~\ref{fig:v-collide} shows $v_t$ and $v_r$ for all collided systems. The black line highlights $v_r^2+v_t^2=v_{\text{esc}}^2$, where $v_{\text{esc}} = \sqrt{2GM_*/R_*}$
is the escape velocity at the radius $r=R_{*,1}+R_{*,2}=2\,R_*$.
Most of the systems lie on this line (as $a_b\gg 2\,R_*$) and the collision speed is roughly given by
\begin{equation}
    v^2=\frac{2GM_*}{R_*}\lrb{1-\frac{R_*}{a_b}}\approx \frac{2GM_*}{R_*}.
\end{equation}
Our results are similar to that obtained by \citet{yu24_tidal_break-up}.
The right panels of Figure~\ref{fig:v-collide} further show the cumulative distribution of $v_r$, assuming a thermal distribution $\eta=1$. Binary collisions occur at a wide range of angles, from nearly radial ($v_r\sim v_{\text{esc}}$) to tangential ($v_r\sim0$). The dependence of the $v_r$ distribution on $a_{b,0}$ is weak, except when the loss cone is empty. In the most empty loss cone regime ($t_{\rm 2B,relax}=10\,{\rm Gyr},a=0.5\,\rm pc$), the collisions are slightly more tangential for large $a_{b,0}$ (see Figure~\ref{fig:delta_l_l_tb}).

\section{Discussion}\label{sec:Discussion}

In this section, we examine the assumptions in our simulations and discuss the broader implications of our results. We first review our use of single oscillation mode ($f$-mode) in modeling the chaotic tides. We then discuss the subsequent consequences of the chaotic tide triggers, including hardening of the inner orbits and the fate of the hardened binaries. Finally, we briefly address the influence of fly-by encounters of field stars on the inner orbits, which is not included in our model.

\subsection{Use of single oscillation mode}\label{sec:Dis-single-mode}
\begin{figure}
    \centering
    \includegraphics[width=\columnwidth]{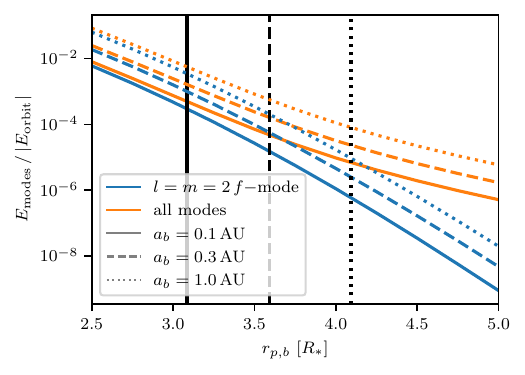}
    \vspace{-0.6cm}
    \caption{The initial energy gain per inner pericenter passage for only the $l=m=2$ $f$-mode (blue lines) and all modes (orange lines). The vertical lines show $r_{b,\rm ct}$ (eq.~\ref{eq:r-ct}). The solid, dashed, dotted lines are for inner SMAs $a_b=0.1,0.3,1.0\,\text{AU}$.}
    \label{fig:E_modes_f_vs_all}
\end{figure}

In our model we only consider the tidal coupling of $l=m=2$ $f$-mode in Star 1 to the inner orbit. Our choice of single oscillation mode is conservative, since the inclusion of other oscillation modes will boost the energy exchange and cause a larger change in the inner orbital period. To estimate the potential modifications from including other oscillation modes, we compute the initial energy kick during the first inner pericenter passage by summing up the energy deposition in Star 1 in all the modes with $l\le 4$ and $0.1<\omega_\alpha/\sqrt{GM_*/R_*^3}<20$:
\begin{align}
    \Delta E_{\text{all modes}}& =\sum_\alpha \Delta E_\alpha = \sum_\alpha \lrb{\Delta A_\alpha}^2,
\end{align}
where $\alpha$ represents each oscillation mode.
Figure~\ref{fig:E_modes_f_vs_all} shows $\Delta E_{f}$ and $\Delta E_{\text{all modes}}$ as functions of $r_{p,b}$ for $a_b=0.1,0.3,1.0\,\text{AU}$. For a given $a_b$, the contribution of $\Delta E_{f\text{-mode}}$ to $\Delta E_{\text{all modes}}$ becomes more and more dominant as $r_{p,b}$ decreases. This is the consequence of two factors: (1) the large tidal overlap integral $Q_\alpha$ of $l=m=2\,f$-mode and (2) the orbital angular frequency near the inner pericenter passage $\Omega_{p,b}$ approaches $\omega_f$ for small $r_{p,b}$. For large $r_{p,b}$, the $g$-modes with lower frequencies can couple to the tidal potential more strongly, and the contribution of the $f$-mode drops. For $r_{p,b}=r_{b,\rm ct}$, the rough pericenter radius where the chaotic growth of $f$-mode is triggered (vertical lines in Figure~\ref{fig:E_modes_f_vs_all}), the contribution of $\Delta E_{f}$ to $\Delta E_{\text{all modes}}$ is $\approx60,40,10\,\%$ for $a_b=0.1,0.3,1.0\,\text{AU}$. Including other oscillation modes will boost energy exchange and increase $f_{\rm CT}$. Nonetheless, the steep dependence of both $\Delta E_{f\text{-mode}}$ and $\Delta E_{\text{all modes}}$ on $r_{p,b}$ suggests that a significant change in the outcome statistics is unlikely.

\subsection{Other stellar masses}
While our simulations only focus on the $0.5\,M_\odot$ MS stars, the physics of stellar oscillation can be applied to stars of other masses as well. For stars with masses $M\gtrsim2\,M_\odot$, the convective core and radiative envelope change the mode structure of stellar oscillations. Nonetheless, \citet{Vick2018} has shown that the transition to chaotic tides also happens with the presence of multiple modes. Due to the steep dependence of the tidal energy injection as the function of $r_{p,b}$ (see Appendix~\ref{sec:App-iter-map} and \ref{sec:temporal-overlapping-integral}), we expect the boundary of chaotic tides to lie at $\text{a few}\times R_*$, regardless of the details of the mode structure.

One important caveat for the more massive stars is their limited lifetime. The binary stars at the Galactic Center undergo outer orbit relaxation on the timescale up to $10^9$ years. The short lifetime of the massive stars means that the fraction of stars that can trigger chaotic tides will be significantly suppressed. A proper account for the stellar evolution \citep[e.g.,][]{Stephan2016,stephan19_eLK} in our treatment of chaotic tides, MBH tidal perturbation, and outer orbit relaxation will be necessary to fully explore the stellar mass dependence, which is outside the scope of our paper.

\subsection{Quenching of chaotic tides}\label{sec:chaotic-tides-damping}

After the triggering of chaotic tides, the $f$-mode energy will grow rapidly and the inner orbit hardens. The growth of the chaotic tides, however, is not indefinite and is instead limited by two factors: (1) the ability of mode-orbit coupling to sustain significant phase change between adjacent orbits, and (2) the damping of $f$-mode energy.

\begin{figure}
    \centering
    \includegraphics[width=\columnwidth]{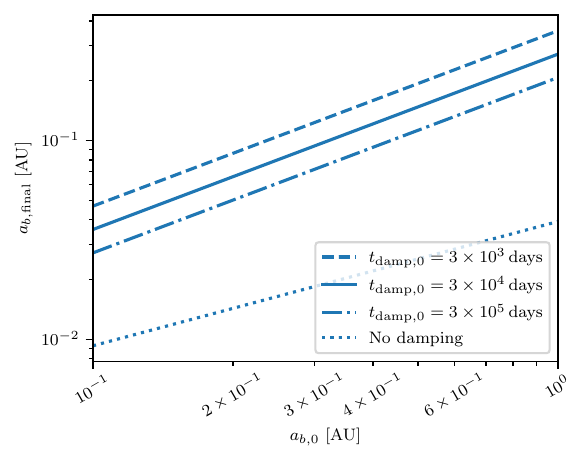}
    \vspace{-0.6cm}
    \caption{The final inner SMA $a_{b,\rm final}$ at the end of chaotic tides, as the function of the initial inner SMA $a_{b,0}$. The chaotic tides are assumed to be triggered marginally. The dotted line corresponds to the idealized situation without damping (eq.~\ref{eq:ab-final-no-damping}). The other lines correspond to the cases where non-linear damping is present (eq.~\ref{eq:ab-final-damping}). We used the non-linear damping prescription from \citet{Kumar1996}, with the variable damping time $t_{\rm damp,0}$ (eq.~\ref{eq:mode-damping-timescale}).
    }
    \label{fig:ab-final}
\end{figure}

We first consider the idealized system without mode damping and that the kick amplitude remains the same as that in the first pericenter passage (for simplicity). The $f$-mode energy evolves with the number of inner orbits $k$ as \citep{Kochanek1992}:
\begin{equation}\label{eq:chaotic-energy-evolution}
    E_{f,k+1} \sim E_{f,k}+2\cos\theta\sqrt{E_{f,k}\Delta E_{f,0}}+\Delta E_{f,0},
\end{equation}
where $\theta$ is the phase difference between the kick and the existing mode. During chaotic tide evolution, the value of $\theta$ can be treated as a uniform random variable in the range $(0,2\pi)$.
Sustaining chaotic tides requires a phase change that is greater than unity:
\begin{equation}
    1<\Delta\phi_k = \omega_f(P_{b,k+1}-P_{b,k}) = \frac{3}{2}\abs{\frac{E_{f,k+1}-E_{f,k}}{E_{b,\rm orb,k}}}\omega_f P_{b,k}.
\end{equation}
If a significant amount of energy is already present in the $f$-mode ($E_{f,k}\approx\abs{E_{b,\text{orb},k}}\gg\abs{E_{b,\text{orb},0}}$), the maximum phase change per orbit is achieved by setting $\cos\theta=1$ in eq.~(\ref{eq:chaotic-energy-evolution}):
\begin{align}\label{eq:chaotic-max-phi-change}
    \max(\Delta\phi_k) &\approx3\abs{\frac{\sqrt{E_{f,k}\Delta E_{f,0}}}{E_{b,\rm orb,k}}} = 3\sqrt{\abs{\frac{\Delta E_{f,0}}{E_{b,\rm orb,k}}}}\omega_f P_{b,k}\nonumber\\
    & = 3\sqrt{\abs{\frac{\Delta E_{f,0}}{E_{b,\rm orb,0}}}}\omega_f P_{b,0}\lrb{\frac{a_{b,k}}{a_{b,0}}}^2.
\end{align}
We conservatively assume that the chaotic tides are initially triggered marginally with no pre-existing mode energy ($\Delta \phi_0=1$ in eq.~\ref{eq:chaotic-max-phi-change}).
The final SMA at the end of chaotic tides $a_{b,\text{final}}$ is determined by setting $\max(\Delta\phi_k)$ to 1 rad:
\begin{equation}\label{eq:ab-final-no-damping}
    a_{b,\text{final}} = a_{b,0}\lrb{6\omega_f P_{b,0}}^{-1/4}\propto a_{b,0}^{5/8},
\end{equation}
which is shown by the dotted line in Figure \ref{fig:ab-final}.

We next consider the more realistic case where the $f$-mode is non-linearly damped when it reaches high amplitudes. 
The non-linear damping timescale in a convective star has been computed by \citet{Kumar1996}:
\begin{equation}\label{eq:mode-damping-timescale}
t_{\text{damp}}=t_{\text{damp,0}}\lrb{\frac{E_f}{10^{42}\,\text{erg}}}^{-1},
\end{equation}
where $t_{\text{damp,0}}$ is estimated to be around $3\times 10^4\,$days. Here we treat it as a variable parameter to account for the fact that the calculations in \citet{Kumar1996} are based on the solar model, not the $0.5\,M_\odot$ binary stars in our simulations.
The incorporation of damping modifies the energy iterative map \citep{Kochanek1992}:
\begin{align}
    E_{f,k+1} = & E_{f,k}\exp\lrsb{-P_{b,k}/t_{\rm damp}(E_{f,k})}+\Delta E_{f,0}\nonumber\\
    &+2\cos\theta\sqrt{E_{f,k}\Delta E_{f,0}}\exp\lrsb{-P_{b,k}/2t_{\rm damp}(E_{f,k})}.
\end{align}
Given the randomness of $\theta$, we model the average $f$-mode energy evolution as the following:
\begin{equation}
    \lara{E_{f,k+1}} = \lara{E_{f,k}}\exp\lrsb{-P_{b,k}/t_{\rm damp}(\lara{E_{f,k}})}+\Delta E_{f,0}.
\end{equation}
The above average iterative model has two evolutionary stages. Initially, the damping of tides is negligible and the average tidal energy grows linearly with the number of inner orbits $\lara{E_{f,k}}\sim k\Delta E_{f,0}$. At later time, the large mode amplitude causes the damping to become more effective. Eventually $\lara{E_{f,k}}$ saturates and the energy kick balances the energy loss from mode damping per inner orbit:
\begin{equation}
    \frac{P_{b,k}}{t_{\text{damp}}(\lara{E_{f,k}})}=\frac{\Delta E_{f,0}}{\lara{E_{f,k}}},
\end{equation}
where we assume that the damping time is long compared to the orbital period, $t_{\text{damp}}(\lara{E_{f,k}})\gg P_{b,k}$.
With the damping timescale prescription in eq.~(\ref{eq:mode-damping-timescale}), the equilibrium $f$-mode energy is given by
\begin{equation}\label{eq:chaotic-tide-equi-E}
    \lara{E_{f,k}}=\lrsb{\frac{t_{\text{damp}}(E=\Delta E_{f,0})}{P_{b,k}}}^{1/2}\Delta E_{f,0}.
\end{equation}
Therefore, in the second stage of chaotic tide growth, the tidal energy will not increase linearly with the number of orbits, but at a smaller rate in response to the shortening of the orbital period. Eventually, the chaotic tides will be quenched when the orbital period change between two consecutive orbits is not sufficient to create enough phase shift in the oscillation mode:
\begin{align}
    1>\max(\Delta\phi_k) = & 3\abs{\frac{\sqrt{\lara{E_{f,k}}\Delta E_{f,0}}}{E_{b,\rm orb,k}}}\omega_f P_{b,k}\nonumber\\
    = & 3\lrsb{\frac{t_{\text{damp}}(E_f=\Delta E_{f,0})}{P_{b,0}}}^{1/4}\nonumber\\
    & \times \lrb{\frac{\Delta E_{f,0}}{E_{b,\rm orb,0}}}\omega_f P_{b,0}\lrb{\frac{a_{b,k}}{a_{b,0}}}^{17/8}.
\end{align}
Again assuming the chaotic tides are marginally triggered, we can obtain $a_{b,\text{final}}$ in the presence of damping:
\begin{equation}\label{eq:ab-final-damping}
    a_{b,\text{final}} \sim a_{b,0}\lrsb{\frac{16\,t_{\text{damp}}(E_f=\Delta E_{f,0})}{P_{b,0}}}^{-2/17} \propto a_{b,0}^{15/17}.
\end{equation}

Figure~\ref{fig:ab-final} shows the final SMA of the binary system $a_{b,\rm final}$ when chaotic tides are quenched.
Four possible cases of tidal damping are considered: no damping, $t_{\text{damp},0}=3\times10^3,3\times10^4,3\times10^5\,$days. In the absence of damping, the chaotic tides have the potential to shrink the inner SMA by a factor of 5--20, depending on $a_{b,0}$. The efficiency of chaotic tides in hardening the inner orbit drops when non-linear damping is present. In the case of very rapid damping ($t_{\text{damp},0}=3\times10^3\,$days), the chaotic tides are only able to shrink the orbit by a factor of 2--3. In general, the chaotic tides have the strongest effect on wide binaries, as $a_{b,\text{final}}\,/\,a_{b,0}\propto a_{b,0}^{-2/17}$ (with damping), $\propto a_{b,0}^{-3/8}$ (without damping).

Note that our analysis above does not take into account the potential structural change in the stellar interior during the chaotic tides. The luminosity of energy injected into the star is on the order of
\begin{equation}
    L_{\text{ct}} \sim \frac{\Delta E_{f,0}}{P_{b,0}} \gtrsim \frac{\abs{E_{b,\text{orb},0}}}{\omega_f P_{b,0}^2} \sim 10^{33}\lrb{\frac{a_{b,0}}{1\,\text{AU}}}^{-4}\,\text{erg}\,\text{s}^{-1}.
\end{equation}
It is unclear if the star can effectively radiate the energy away. It is possible that the star will expand in radius and the tidal interaction strengthens \citep{Kumar1996}, potentially leading to either a runaway or mass loss from the stellar surface.

\subsection{Evolution of binaries after chaotic tides}\label{sec:after-chaotic-tides}
At the end of chaotic tides, the binary systems have $r_{p,b}\sim2-4\,R_*$ and $a_b\sim{\text{a few}}\times10\,R_*$ depending on the initial inner SMA $a_{b,0}$ and the efficiency of non-linear damping (Figure~\ref{fig:ab-final}). The subsequent evolution of those binaries is governed by two effects: (1) the further circularization of the inner orbits due to the (non-chaotic) tidal interactions between binary stars, and (2) the outer orbit angular momentum relaxation. In this subsection, we discuss the importance of tidal circularization and the final outcome of the binaries.

Studies on the circularization of close binaries have found that dynamical tides can in general provide much more efficient damping compared to equilibrium tides at short periods \citep{Goodman1998,Barker2020}.
Those studies mainly concern the systems with small eccentricities and are not directly applicable to binaries at the end of chaotic tides, which still possess significant eccentricity (especially the cases with non-linear damping).
In Appendix~\ref{sec:App-dyn-tides}, we extrapolate their results to higher eccentricities based on physical arguments. Despite the great uncertainties associated with the orbital circularization timescale, our estimate suggests that dynamical tides have the potential to harden binaries to $a_b\lesssim10\,R_*$ within one outer orbit, which is much faster than the angular momentum relaxation.

Rapid hardening of binaries leads to a significant shrinkage of $r_t$, and can save them from the originally imminent Hills breakup. If the binaries later drift away from the MBH due to relaxation, they have the potential to become eclipsing binaries when observed nearly edge-on.
On the other hand, if further relaxation brings the hardened binary closer to the MBH and causes it to undergo Hills breakup, the bound star will be delivered on a tighter orbit, with
\begin{align}
    r_p &\sim a_b \lrb{\frac{\MBH}{M_b}}^{1/3}\sim4\,{\rm AU}\lrb{\frac{a_b}{10\,R_*}},\\
    a &\gtrsim a_b \lrb{\frac{\MBH}{M_b}}^{2/3}\sim6\times10^2\,{\rm AU}\lrb{\frac{a_b}{10\,R_*}},\\
    P& = 2\pi\sqrt{\frac{a^3}{G\MBH}}\gtrsim 7\,{\rm yr}\lrb{\frac{a_b}{10\,R_*}}^{3/2}.
\end{align}
The Hills breakup of tight binaries may thus be a major source of repeating partial TDEs that have periods on the order of months to years \citep{Cufari2022}.
Furthermore, the steep dependence of $t_{\rm GW}$ on $r_p$ means that the bound stars from those tightest binaries will have the highest chance of avoiding TDE and becoming EMRI through GW circularization (Figure~\ref{fig:bound-star}):
\begin{align}
    \frac{t_{\text{GW}}}{t_{L,\text{relax}}} \propto r_p^{5/2}a^{3/2} t_{\text{2B,relax}}^{-1}\propto a_b^4.
\end{align}
These EMRI may be required to explain the QPEs discovered in recent years \citep{linial23_QPE_EMRI_TDE,lu23_QPE_from_EMRI}.
The tight binaries also have the potential to produce the fastest hyper-velocity stars and can explain the high velocity of S5-HVS1 \citep{Koposov2020}.

\subsection{Influence of other field stars}
In our model, we consider the outer orbital relaxation due to the weak gravitational encounters between the binary system and the field stars.
In principle, those weak encounters can also influence the inner orbit of binary stars.
In the following, we evaluate the significance of those effects and discuss their implications.

For relatively wide binaries ($a_b\ge 0.1\,{\rm AU}$), the inner orbital velocities are much smaller than the velocity dispersion in the nuclear star cluster. Over time, the interaction with other field stars tends to increase the inner orbital energy and evaporate the binaries \citep{Heggie1975}.
For a binary system surrounded by field stars with velocity dispersion $\sigma\sim\sqrt{G\MBH/r}$ and stellar number density $n$, the local evaporation timescale is roughly \citep{Binney2008}
\begin{equation}
    t_{\rm evap}(r)\sim 0.06\frac{\sigma}{GM_{\odot}na_b\ln\Lambda},
\end{equation}
where we assume field stars of $1\,M_\odot$ and ignore the order-of-unity dependence on the binary stellar mass. The Coulomb logarithm $\ln\Lambda$ is determined by
\begin{equation}
    \ln\Lambda=\ln\lrb{\frac{a_b / 2}{R_*}}.
\end{equation}
We take $\ln\Lambda\sim 5$ for our wide binaries ($a_b\sim 0.3\,\rm AU$).
For a power law density profile that is not too steep ($n\propto r^{-\gamma};\gamma<2$), the interactions near the outer apocenter dominate the contribution to evaporation. Using the stellar mass density $\rho(r) = n(r) M_\odot$ in eq.~(\ref{eq:density-profile}), $t_{\rm evap}$ is estimated to be:
\begin{equation}
    t_{\rm evap}(r=a) \sim 10^9\,{\rm yr}\lrb{\frac{a}{1\,\rm pc}}^{1.25}\lrb{\frac{a_b}{0.3\,\rm AU}}^{-1}.
\end{equation}
This is in general longer than $t_{L,\rm relax}$ for binaries in our model (eq.~\ref{eq:t-relax-AM}):
\begin{equation}
    t_{L,\rm relax} \sim 3\times10^7\,{\rm yr}\lrb{\frac{a}{1\,\rm pc}}^{-1}\lrb{\frac{a_b}{0.3\,\rm AU}}\lrb{\frac{t_{\rm 2B,relax}}{1\,\rm Gyr}}.
\end{equation}
where we use the initial outer pericenter $r_{p,0}=70\,r_t$.
The influence of evaporation on our system is therefore limited, except for the widest binaries ($a_b=1.0\rm AU$) under inefficient relaxation ($t_{\rm 2B,relax}=10\,\rm Gyr$). We note, however, that the above analysis is only applicable to binary systems born on eccentric outer orbits. For binary systems on less eccentric outer orbits, they have much larger $t_{L,\rm relax}$ and evaporation can be significant to their evolution.

Another possible outcome of fly-by encounters is the collision between the field stars and the binary system. In general, the collision timescale is not sensitive to the eccentricity of the outer orbit due to gravitational focusing at large radii \citep{Rose2020}, and is on the order of \citep{Binney2008}
\begin{equation}
    t_{\rm col}\sim\frac{1}{16\sqrt{\pi}n\sigma(2R_*)^2}\lrb{1+\frac{GM_*/R_*}{4\sigma^2}}^{-1}.
\end{equation}
For $a=1\,\rm pc$, $t_{\rm col}\sim 10^{11}\,{\rm yr}\gg t_{L,\rm relax}$ and the collision probability is therefore negligible for our binaries with large outer orbits\footnote{Note that the effects of collision can be significant for $a\ll1\,\rm pc$ \citep[see e.g.,][]{rose23_stellar_collisions}.}.

Lastly, the gravitational encounters with field stars can also perturb the inner orbital angular momentum and change $e_b$. This can happen on a timescale much less than $t_{\rm evap}$ for highly eccentric inner orbits, as the small orbital angular momentum makes the orbit more susceptible to perturbations. Properly evaluating the effects of gravitational encounters on $e_b$ is outside the scope of this paper. However, recently \citet{Winter-Granic2024} found that the perturbation on $e_b$ from stellar flybys, combined with the tidal field of the nuclear star cluster, can drive the $e_b$ to extremely high values. This may offer a new venue for triggering the chaotic tides and the formation of hard binaries at the Galactic Center \citep[e.g.,][]{dodici25_chaotic_tides_KL_and_flyby}.

\subsection{Fraction of chaotic tides}
Our simulations explore the dependence of various binary outcome fractions on the emptiness of the loss cone.
Extending these results to the broader Galactic Center binary population requires understanding of the binary fraction and the distribution of inner orbital elements, both of which are shaped by stellar flybys through evaporation, hardening, and eccentricity perturbations \citep{Winter-Granic2024}.
Furthermore, our results specifically describe binaries that have relaxed into highly eccentric outer orbits and experience strong MBH tidal perturbations; these represent only a subset of the total population born on more circular orbits.
The fraction of binaries excited to these high eccentricities is sensitive to the uncertain relaxation time profile, which also determines the emptiness of the loss cone --- a key parameter in the onset of chaotic tides.
For binaries with outer SMA near or beyond the MBH's sphere of influence, collisionless relaxation may further modify the framework assumed here \citep{Penoyre2025}, particularly if its timescale is shorter than that of traditional two-body relaxation.
Alternatively, chaotic tides may be triggered by weaker MBH perturbations through eZLK oscillations \citep{dodici25_chaotic_tides_KL_and_flyby}.
Future observations will be essential to constraining the binary population at the Galactic Center.

\section{Conclusions}\label{sec:Conclusions}

In this paper we study the evolution of binary systems orbiting a MBH (taking our own Galactic Center as an example).
Our model includes the influence of tidal perturbations from the MBH, gravitational relaxation of the outer orbit, and tidal interactions between the two stars (Figure~\ref{fig:system-overview}).
Our work is motivated by the fact that, unlike the classical Hills mechanism where binary systems approach the MBH and are tidally broken apart in one outer pericenter passage \citep{Hills1988}, the inefficient relaxation (empty loss cone) causes the outer pericenter radius $r_{p}$ to evolve diffusively, and it takes a large number of outer orbits of relaxation before the binary can reach the disruption radius $r_t$. During the evolution, the MBH's tides repeatedly perturb the binary's inner eccentricity $e_b$.
We simulate the perturbation process through analytical prescriptions of secular approximation \citep{Hamers2019} and REBOUND \citep{Rein2012,Rein2015}.
Our results show that a significant population of binary systems can become highly eccentric, with $r_{p,b}\lesssim \text{a few}\times R_*$, before Hills breakups and stellar collisions.
At such small inner pericenter radii, the tidal interactions between binary stars excite the stellar oscillation modes chaotically and lead to a diffusive growth of oscillation amplitudes (Figure~\ref{fig:chaotic-tides-example}).
Based on analytical arguments and the ``iterative map'' model \citep{Vick2018}, we show that the chaotic tides are highly efficient and can harden the binaries within one outer orbit ($\sim 10^5\,{\rm yr}$).

Our detailed Monte Carlo simulations further show that the wide binaries ($a_b \sim 1.0\,{\rm AU}$) undergoing inefficient outer orbit relaxation ($t_{\rm 2B,relax} \sim 10\,{\rm Gyr}$) are the most susceptible to chaotic tides, and up to $50\%$ of the systems that approach the MBH closely can trigger chaotic tides (Figure~\ref{fig:outcome-fractions}).
Even though their subsequent evolution is not modeled in detail, we expect that a combination of non-linear damping of the chaotically excited tides (\S\ref{sec:chaotic-tides-damping}) and finite-amplitude dynamical tides (\S\ref{sec:after-chaotic-tides}) can further shrink binaries to $a_b\lesssim 10\,R_*$.
The evolution of these tight binaries is of great interest. Hills breakup may eject one binary star as an extreme HVS, potentially explaining the high velocity of S5-HVS1 \citep{Koposov2020}. On the other hand, the bound companion could evolve into EMRIs through GW orbital decay and produce repeating nuclear transients such as partial TDEs and QPEs \citep{Cufari2022,Payne2021,Miniutti2019,Linial2023, lu23_QPE_from_EMRI, linial23_QPE_EMRI_TDE, yao25_tidal_heating}.

In our model we only consider one stellar oscillation mode ($l=m=2$ $f$-mode). In principle other oscillation modes also contribute to the tidal interaction, and damping of the dynamical tides of those oscillation modes may help harden the binaries even before the onset of the f-mode chaotic tides. Furthermore, gravitational encounters with other field stars may also perturb the inner eccentricity and lead to tidal dissipation via the eZLK mechanism \citep{Winter-Granic2024, dodici25_chaotic_tides_KL_and_flyby}. Both of these effects may contribute to the formation of close binaries at galactic centers.

In summary, our work extends the traditional picture of Hills breakup to include the long-term evolution of binary systems. Through MBH tidal perturbations and tidal interactions between binary stars, we demonstrate the effectiveness of chaotic tides at hardening the binaries. A significant population of close binaries formed through this mechanism may be present at our own Galactic Center and in the nuclei of other galaxies. Future works on their evolution will be essential to uncovering the connection between such binaries and repeating partial TDEs or QPEs.

\section*{Acknowledgments}

We thank Dong Lai for helpful suggestions along this project. The research of HH and WL are supported by Rose Hills Innovator Program. We thank the participants of the ZTF Theory Network meeting at Oak Creek, especially Jim Fuller, Itai Linial, Eliot Quataert, and Sterl Phinney, for many stimulating discussions and this research benefited from interactions supported by the Gordon and Betty Moore Foundation through Grant GBMF5076. We also thank the organizers (especially Giovanni Miniutti) of the X-ray Quasi-Periodic Eruptions \& Repeating Nuclear Transients Conference in Spain where we have received helpful comments on this work.

\bibliographystyle{mnras}

\bibliography{refs}

\begin{appendix}

\section{Precessions of the inner and outer orbits}\label{sec:App-precessions}

There are several precession mechanisms in the MBH-binary system, the leading-order ones being the Schwarzschild precession of the inner and outer orbits, and the mass precession of the outer orbit. The precessions mainly affect the relative orientation between the inner and outer orbits, and influence the MBH perturbation on the binary (Section~\ref{sec:Methods-MBH-perturbation}). In this subsection we evaluate the importance of those precessions in our modeling.

\subsection{Schwarzschild precession of inner/outer orbits}
The Schwarzschild precession per inner orbit is given by
\begin{equation}
    \Delta \omega_{s,\text{inner}} = \frac{6\pi GM_{b}}{c^{2}a_{b}(1-e_{b}^{2})}.
\end{equation}
Since $r_{p,b}\gtrsim r_{b,\rm ct}$ before the chaotic tides, $\Delta \omega_{s,\text{inner}}$ is limited by:
\begin{equation}
    \Delta \omega_{s,\text{inner}} \lesssim \frac{3\pi GM_{\text{b}}}{c^{2}r_{b,\rm ct}} \simeq 1\times 10^{-5}\lrb{\frac{r_{b,\rm ct}}{4\,R_*}}^{-1}.
\end{equation}
Despite the small value of $\Delta \omega_{s,\text{inner}}$, the cumulative inner orbital Schwarzschild precession over one outer orbit can be significant:
\begin{align}
    \Delta \omega_{s,\text{inner,tot}} &= \Delta \omega_{s,\text{inner}}\lrb{\frac{P}{P_b}} \nonumber\\
    &\lesssim 0.48\lrb{\frac{a}{1\,\text{pc}}}^{3/2}\lrb{\frac{a_b}{1\,\text{AU}}}^{-3/2}\lrb{\frac{r_{b,\rm ct}}{4\,R_*}}^{-1}.
\end{align}
The inner orbit Schwarzschild precession directly changes the result of each MBH perturbation, which can be seen from the first-order perturbation theory \citep{Heggie1996}.
We include the inner orbit Schwarzschild precession in our model by adding $\Delta \omega_{s,\text{inner,tot}}$ to the inner argument of pericenter before the MBH perturbation every outer orbit.

While the outer orbit also undergoes Schwarzschild precession, it is much weaker:
\begin{equation}
    \Delta \omega_{s,\text{outer}} \approx \frac{3\pi G\MBH}{c^{2}r_p} = 3.6\times10^{-5}\lrb{\frac{r_p}{70\,r_t}}^{-1}\lrb{\frac{a_b}{1\,\text{AU}}}^{-1}.
\end{equation}
More specifically, the outer orbital Schwarzschild precession is slow compared to the angular momentum relaxation
\begin{align}
    \Delta \omega_{s,\text{outer}}N_l &\approx \frac{6\pi G\MBH}{c^{2}a}\frac{t_{\text{2B,relax}}}{P(a)}\nonumber\\
    &= 0.08\lrb{\frac{t_{\text{2B,relax}}}{1\,\text{Gyr}}}\lrb{\frac{a}{1\,\text{pc}}}^{-5/2}.
\end{align}
Since the orientation of the outer orbit is expected to change significantly due to mass precession (see below) even without the Schwarzschild precession, $\Delta \omega_{s,\text{outer}}$ is unimportant and hence not included in our model.

\subsection{Mass precession of outer orbit}
Due to the extended distribution of the stellar/compact object population near the MBH, the outer orbit is subject to the mass precession. Ignoring the order-of-unity correction from the density profile, the mass precession per outer orbit is given by \citep{Merritt2013}
\begin{align}
    \Delta\omega_{\text{mass}} \sim -2\pi\sqrt{1-e^2}\frac{M(a)}{\MBH}\sim -2\pi\sqrt{\frac{2r_p}{a}}\frac{M(a)}{\MBH},
\end{align}
where $M(a)$ is the extended mass within radius $a$ from the MBH.
We use the density profile from \citet{Schodel2007}
\begin{equation}\label{eq:density-profile}
    \rho(r)=2.8\times10^6\,M_\odot\,\text{pc}^{-3}\times\lrb{\frac{r}{0.22\,\text{pc}}}^{-\gamma},
\end{equation}
where $\gamma=1.2$ for $r<0.22\,\text{pc}$ and $\gamma=1.75$ for $r>0.22\,\text{pc}$. The corresponding $M(r)$ for $r>0.22\,\rm pc$ is
\begin{equation}
    M(r)= 3.0\times10^5\,M_\odot\lrb{\frac{r}{0.22\,\text{pc}}}^{1.25}-9\times 10^4\,M_\odot.
\end{equation}
For a binary system with $a_b=70\,r_t$ and $a=1.0\,\rm pc$, $M(r=a)\sim2\times10^6\,M_\odot$ and $\Delta\omega_{\text{mass}}\sim -1\rm\, rad$. Even though the mass precession does not directly affect the MBH perturbation on the binary, its indirectly influence orientation and the relaxation of the outer orbit.
In our model, the mass precession is included by adding $\Delta\omega_{\text{mass}}$ to the outer argument of pericenter before the MBH perturbation every outer orbit.

\section{Method of iterative map}\label{sec:App-iter-map}
(In this section the subscript $_b$ is omitted, and all orbital properties refer to the inner orbit.)

The iterative map concerns the stellar/planet oscillation mode evolution in highly eccentric binary systems.
It was first introduced by \citet{Ivanov2004} and further developed by \citet{Wu2018, Vick2018}. Here we briefly summarize \citet{Vick2018}'s approach.
For a non-rotating star (Star 1), the general first-order Lagrangian displacement of the fluid elements $\xi({\bf x},t)$ can be expanded with the eigenmodes $\alpha$ with eigenfunctions $\xi_\alpha({\bf x})$ and eigenfrequencies $\omega_\alpha$ \citep{Schenk2001}:
\begin{align}
    \begin{bmatrix}
        \boldsymbol{\xi}({\bf x},t) \\ \frac{\partial\boldsymbol{\xi}({\bf x},t)}{\partial t}
    \end{bmatrix}
    = \sum_\alpha c_\alpha(t)
    \begin{bmatrix}
        \boldsymbol{\xi}_\alpha({\bf x}) \\ -i\omega_\alpha\boldsymbol{\xi}_\alpha({\bf x})
    \end{bmatrix}.
\end{align}
The eigenfunctions are normalized with $\int_V\rho \boldsymbol{\xi}_\alpha^*\cdot\boldsymbol{\xi}_\alpha d^3x=1$.
In the presence of an external tidal potential from the companion (Star 2), the coefficients $c_\alpha$ evolve according to
\begin{align}
    \dot{c}_\alpha+i\omega_\alpha c_\alpha = \frac{i}{2\omega_\alpha}\frac{GM_{b,2}}{r(t)^{l+1}}W_{lm}Q_\alpha e^{-im\Phi(t)},\label{eq:App-c-ode}
\end{align}
where $r(t)$ is the binary separation and $\Phi(t)$ is the true anomaly of the orbit. The damping is ignored in the above expression. $W_{lm}$ is a numerical constant and $Q_\alpha$ is the tidal overlap integral that depends on the spatial structure of the oscillation modes \citep{Press1977}:
\begin{align}
    Q_\alpha = \int_V\rho\boldsymbol{\xi}_\alpha^*\cdot\nabla\lrb{r^l Y_{lm}}d^3x,
\end{align}
where $\rho$ is the stellar density profile prior to perturbation and $Y_{lm}$ is the spherical harmonics corresponding to the mode $\alpha$.
Another equivalent expression of $Q_\alpha$ is \citep{Burkart2012,Fuller2017}
\begin{align}
    Q_\alpha = -(2l+1)\frac{R_{*,1}^{l+1}}{4\pi G}\tilde{\Phi}_\alpha (R_{*,1}),
\end{align}
where $\tilde{\Phi}_\alpha (R_{*,1})$ is the Eulerian gravitational potential perturbation on the surface of the star.
We further define the dimensionless tidal overlap integral $\bar{Q}_\alpha$:
\begin{align}
    \bar{Q}_\alpha & = \frac{Q_\alpha}{M_{*,1}^{1/2}R_{*,1}^{l-1}}.
\end{align}

While in principle eq.~(\ref{eq:App-c-ode}) can combined with the back reaction of the oscillation on the orbital dynamics to solve the tidal evolution of the binary system \citep{Wu2018}, the drastically different timescales between orbital dynamics and stellar oscillations make numerical integrations very challenging. Iterative maps utilize the fact that in highly eccentric orbits, the tidal interaction and driving of oscillation modes are limited to the region near pericenter. By approximating the orbits between consecutive pericenter passages as ellipses, the stellar oscillations can be solved iteratively over multiple orbits. We focus on the iterative map of only $l=m=2$ $f$-mode, which dominates the oscillation energy in a highly eccentric binary of low mass stars (see Section~\ref{sec:Dis-single-mode}). For $0.5\,M_\odot$ ZAMS stars, this mode has $\omega_f/\Omega_{*,1}=1.67084$ and $\bar{Q}_f = 0.1198$ based on MESA and GYRE calculations. Here $\Omega_{*,1}=\sqrt{GM_{*,1}/R_{*,1}^3}$ is the angular frequency associated with the dynamical time of Star 1.

Let $t_k$ be the time of $k$-th apocenter passage, $t_{k+1/2}$ be the time of $k$-th pericenter passage, and $c_{f,k}=c_f(t_k)$. Define a new variable $A_{f,k}$ to represent the amplitude of the oscillation modes:
\begin{equation}
    A_{f,k} = \sqrt{2}\omega_f c_{f,k}e^{-i\omega_f P_{k}/2},
\end{equation}
where $P_{k}=t_{k+1/2}-t_{k-1/2}$ is the period of the $k$-th inner orbit. The energy and angular momentum in the mode are $E_{f,k}=\abs{A_{f,k}}^2,L_{f,k}=m\abs{A_{f,k}}^2/\omega_f$. The evolution of $A_{f,k}$ follows the iterative map below:
\begin{equation}
    A_{f,k+1} = \lrb{A_{f,k+1}+\Delta A_{f,k}}e^{i\omega_f P_{k+1}}.
\end{equation}
The energy transfer due to the tidal interaction during the pericenter passage is encapsulated in $\Delta A_k$:
\begin{align}
    \Delta A_{f,k} = \frac{i\sqrt{2}\pi GM_{*,2}Q_f}{r_{p,k}^{l+1}}\frac{1}{\Omega_{*,1}}K_{f,k},
\end{align}
where $r_{p,k}$ is the $k$-th pericenter radius, and $K_{f,k}$ is an integral that quantifies the temporal coupling between the oscillation mode and the tidal potential. In general, $K_{f}$ depends on $\omega_f$ and orbital parameters:
\begin{align}\label{eq:K_integral}
    K_{f} = \frac{W_{lm}}{2\pi}\Omega_{*,1}\int_{-P/2}^{P/2}\lrsb{\frac{r_{p}}{r(t)}}^{l+1}e^{i\lrsb{\omega_f t^\prime-m\Phi(t^\prime)}}dt^\prime,
\end{align}
where the integration is performed on an elliptical orbit with SMA $a$ and eccentricity $e$. Due to the quick oscillations and the large value of $\omega_f$, numerically evaluating $K_{f}$ is computationally expensive.
\citet{Lai1997} provided an analytical approximation for $l=m=2$ mode in the parabolic orbit limit where $\omega_f$ is much larger than the angular frequency of the orbit at the pericenter. In Appendix~\ref{sec:temporal-overlapping-integral}, we generalize their results to elliptical orbits with high eccentricities.

To compute the values of $P_k,K_{f,k}, \Delta A_{f,k}$ and complete the iterative map, one must know $a_k$ and $e_k$, the SMA and the eccentricity between $(k-1)$-th and $k$-th pericenter passages. Without oscillation damping, $a_k$ and $e_k$ are determined by the conservation of the total energy and angular momentum:
\begin{align}
    E_{\text{total}} & = E_{\text{orb},k} + E_f  = -\frac{GM_b\mu}{2a_k}+\abs{A_{f,k}}^2,\\
    L_{\text{total}} & = L_{\text{orb},k} + L_f  = \mu\sqrt{GM_b a_b(1-e_k^2)}+\frac{m}{\omega_f}\abs{A_{f,k}}^2,
\end{align}
where $\mu=M_{*,1}M_{*,2}/(M_{*,1}+M_{*,2})$ is the reduced mass of the binary system.

Without pre-existing oscillations, the first inner pericenter passage injects energy $\Delta E_{f,0}$ into $f$-mode:
\begin{equation}
    \Delta E_{f,0} = 2\pi^2\frac{GM_{*,1}^2}{R_{*,1}}\lrb{\frac{M_{*,2}}{M_{*,1}}}^{2}\lrb{\frac{R_{*,1}}{r_p}}^{2(l+1)}\bar{Q}_{f}^2K_f^2,
\end{equation}
which is a very steep function of $r_p$ due to the $2(l+1)$ power law and the non-linear exponential nature of the temporal overlapping integral $K_f$.
The energy injection in the oscillation mode will cause a change in the orbital period:
\begin{align}
    \frac{\Delta P}{P} & \approx -\frac{3}{2}\frac{\Delta E_{\text{orb}}}{E_{\text{orb}}} = \frac{3}{2}\frac{\Delta E_{f,0}}{E_{\text{orb}}}.
\end{align}
When the period change is large compared to $\omega_f$ ($\Delta\phi=\omega_f\Delta P\gtrsim 1$), the next pericenter passage will introduce a kick in $A_f$ at an effectively random phase. The repeated kicks at random phases over many pericenter passages will result in diffusive and chaotic growth of $A_f$ and $E_f$.
The average energy growth rate is
\begin{equation}
    E_f\sim N\Delta E_{f,0},
\end{equation}
where $N$ is the number of inner orbits.

\section{Analytic approximation of the temporal overlapping integral for elliptical orbits}\label{sec:temporal-overlapping-integral}

Our goal is to analytically compute temporal overlapping integral $K$ (hereafter removing the mode-identity index $f$) in eq.~(\ref{eq:K_integral}) at the quadrupolar order $l=2$ and for the energetically dominating prograde mode with $m=2$. We consider an elliptical orbit with eccentricity $e<1$. For a high eccentricity $e\approx 1$, it is convenient to define a small quantity $\epsilon\ll 1$ as follows
\begin{equation}
    \epsilon \equiv {1-e\over 1+e} \ \ \Leftrightarrow\ \ e\equiv {1-\epsilon\over 1+\epsilon}.
\end{equation}
We define a convenient integration variable $z$ based on the true anomaly $\Phi$ of the orbit,
\begin{equation}
    z\equiv \tan (\Phi/2), \ \cos\Phi = {1-z^2\over 1+z^2},\  \sin\Phi = {2z\over 1+z^2},\ \d\Phi = {2\d z\over 1+z^2},
\end{equation}
and an elliptical orbit with eccentricity $e$ and pericenter radius $r_p$ can be described by
\begin{equation}
    {r\over r_p} = {1+e\over 1 + e\cos\Phi}= {1+z^2\over 1+\epsilon z^2},
\end{equation}
and
\begin{equation}
    t(z) = {1\over \Omega_p} \int_0^\Phi {r^2\over r_p^2} \d \Phi =  {2\over \Omega_p} \int_0^z  {1+z^2\over (1+\epsilon z^2)^2} \d z,
\end{equation}
where $\Omega_p = L/r_p^2$ is the angular frequency at the pericenter for specific orbital angular momentum $L$. The key difficulty in carrying out the time integral in $K$ is the phase factor $\omega t(z)$. Fortunately, one can use the saddle point method to evaluate this integral as long as $\omega\gg \Omega_p$, which is the case in our consideration.

Let us define a dimensionless quantity $$\lambda \equiv 2\omega/\Omega_p,$$ and the mode oscillation phase factor $i\omega t(z)$ can be described by defining a complex function $h(z)$ as follows
\begin{equation}
\begin{split}
    i\omega t(z) &\equiv \lambda h(z),\\
    h(z)
    &= {i\over 2\epsilon^{3/2}} \lrsb{(1+\epsilon)\tan^{-1}(\sqrt{\epsilon} z) - (1-\epsilon) {\sqrt{\epsilon} z\over 1 + \epsilon z^2}}. \\
\end{split}
\end{equation}

For a given mode frequency $\omega$, the temporal overlapping integral $K$ for $l=m=2$ can be written as
\begin{equation}\label{eq:K_integral_gh}
\begin{split}
    {\pi \over W_{22}}{\Omega_{p}\over \Omega_{*,1}} K &= {1\over 2} \int_{-\pi}^{\pi} {r_{p} \over r(\Phi)} \mathrm{e}^{i\lrsb{\omega t(\Phi) + 2 \Phi}} \d \Phi\\
    &= \int_{-\infty}^{+\infty} {1+\epsilon z^2\over 1+z^2} \lrb{i-z\over i+z}^2 \mathrm{e}^{i\omega t(z)} {\d z\over 1+z^2}\\
    &= \int_{-\infty}^{\infty} g(z)\, \mathrm{e}^{\lambda h(z)} \d z,
\end{split}
\end{equation}
where we have made use of
\begin{equation}
    \mathrm{e}^{-2i\Phi} = \lrb{\cos\Phi - i\sin\Phi}^2 = \lrsb{(i+z)/(i-z)}^2,
\end{equation}
and defined the complex function $g(z)$,
\begin{equation}
    g(z) = {1+\epsilon z^2\over (1+z^2)^2} \lrb{i+z \over i-z}^2 = {1 + \epsilon z^2\over (z-i)^4}.
\end{equation}

Although $z$ is a real physical quantity as it is based on the true anomaly $\Phi$, we consider the final expression in eq.~(\ref{eq:K_integral_gh}) to be a complex path integral with $z = x + iy$ which is carried out along the real ($x$) axis of the complex plane --- the original path goes from point $A(x=-\infty, y=0)$ to point $B(x=\infty, y=0)$. According to Cauchy's residue theorem, we may choose an alternative (arbitrary) path that connects $A$ and $B$, and the results will be the same as long as the loop formed between the original path (along the real axis) and the alternative path does not enclose any poles of the integrand function $g(z)\mathrm{e}^{\lambda h(z)}$.

In the following, we will choose the path where the argument in the exponential term $\mathrm{e}^{\lambda h(z)}$ has the steepest descent near a saddle point. When $\lambda\gg 1$ and $h = u(z) + iv(z)$, most of the contribution to the above integral should come from the region where $u(z)$ is maximized while $v(z)$ stays constant (to avoid cancellation due to oscillations). This region must be near where the first derivative $h'(z)=0$, the solutions to which are the \textit{saddle points} of the function $h(z)$.

Since $h'(z) = (1+z^2)/(1+\epsilon z^2)^2$, we find two saddle points at $z_0=\pm i$. We will choose $z_0=i$ as $\mathrm{Re}[h''(z_0)] > 0$ and hence the integral can be converted into the Gaussian form. The 2nd derivative of the function $h(z)$ is given by
\begin{equation}
    h''(z) = 2iz {1 - 2\epsilon - \epsilon z^2\over (1 + \epsilon z^2)^3}
\end{equation}
and we obtain
\begin{equation}
    h''(z_0=i) = - {2\over (1-\epsilon)^2},
\end{equation}
which is indeed negative. Similarly, we carry out the third derivative and evaluate it at the saddle point $z_0=i$,
\begin{equation}
    h'''(z_0=i) = 2i {1 + 7\epsilon\over (1-\epsilon)^3}.
\end{equation}
We can also evaluate the original function $h(z)$ at the saddle point in the limit of $\epsilon\ll 1$
\begin{equation}
    h(z_0=i) \approx -{2\over 3}\lrb{1 + {2\over 5}\epsilon + {9\over 35}\epsilon^2 + {4\over 21}\epsilon^3 }.
\end{equation}
The above results allow us to Taylor expand $h(z)$ around $z_0=i$,
\begin{equation}
\begin{split}
    h(z) &\approx h(z_0) + {h''(z_0)\over 2} (z - z_0)^2 + {h'''(z_0)\over 6} (z - z_0)^3\\
    &\approx -h_0 - h_2 \tilde{z}^2 + i h_3 \tilde{z}^3,
\end{split}
\end{equation}
where
\begin{equation}
    \tilde{z} \equiv z-z_0,
\end{equation}
and
\begin{equation}
\begin{split}
    h_0 &\approx {2\over 3}\lrb{1 + {2\over 5}\epsilon + {9\over 35}\epsilon^2 + {4\over 21}\epsilon^3},\\
    h_2 &= {1\over (1-\epsilon)^2}, \\
    h_3 &= {{1\over 3} {1 + 7 \epsilon \over (1-\epsilon)^3}}.
\end{split}
\end{equation}
Let us also re-write the function $g(z)$ in terms of $\tilde{z} = z-z_0$,
\begin{equation}
    g(z) = \tilde{z}^{-4} \lrsb{1 + \epsilon (\tilde{z} + i)^2} = \sum_{n=0,1,2} g_n i^n \tilde{z}^{n-4},
\end{equation}
where
\begin{equation}\label{eq:gn012}
    g_0 = 1 - \epsilon, \ \ 
    g_1 = 2\epsilon, \ \ 
    g_2 = -\epsilon.
\end{equation}
Thus, the integral in eq.~(\ref{eq:K_integral_gh}) can be written as
\begin{equation}
\begin{split}
    I(\lambda, \epsilon) &= \int_{-\infty}^{\infty} g(z)\, \mathrm{e}^{\lambda h(z)} \d z\\
    &\approx \mathrm{e}^{-\lambda h_0} \int_{\mathcal{C}} \lrb{\sum_{n=0,1,2} g_n i^n \tilde{z}^{n-4}} \mathrm{e}^{-\lambda h_2 \tilde{z}^2} \mathrm{e}^{i\lambda h_3 \tilde{z}^3}\d z,
\end{split}
\end{equation}
where the path $\mathcal{C}$ is determined as follows.

\begin{figure}
    \centering
    \includegraphics[width=0.5\columnwidth]{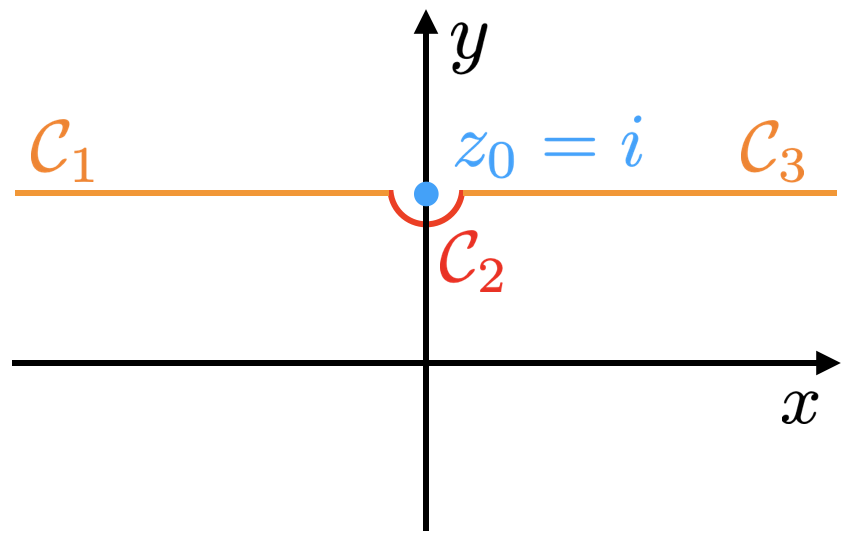}
    \caption{The complex integral path $\mathcal{C} = \mathcal{C}_1 + \mathcal{C}_2 + \mathcal{C}_3$ that passes close to the saddle point $z_0=i$.
    }
    \label{fig:integral_path}
\end{figure}

To avoid rapid oscillations in the Gaussian-like term $\mathrm{e}^{-\lambda h_2 \tilde{z}^2}$, we would like to take the path $\mathcal{C}$ to be perpendicular to the imaginary axis. Looking at the polynomial expansion above, we see that the integrand has a pole exactly at the saddle point at $z_0=i$. For this reason, we slightly change the path into three segments that go around the pole: $\mathcal{C}=\mathcal{C}_1 + \mathcal{C}_2 + \mathcal{C}_3$ as shown in Fig. \ref{fig:integral_path}, where $C_1$ is a straight line from $-\infty+i$ to $-\rho +i$, $C_2$ is a semicircle of infinitesimal radius $\rho$ from $-\rho+i$ to $\rho+i$, and $C_3$ is a straight line from $\rho+i$ to $+\infty + i$. The semicircle is taken to be below $z_0=i$ so the residue at this pole is not involved in the final result. Note that, although the end points of $-\infty + i$ and  $+\infty + i$ do not lie on the real axis, this makes little difference as long as most of the contribution to the integral comes from the region near the saddle point $z_0=i$ --- this is the case for highly eccentric orbits.

\begin{figure}
    \centering
    \includegraphics[width=\columnwidth]{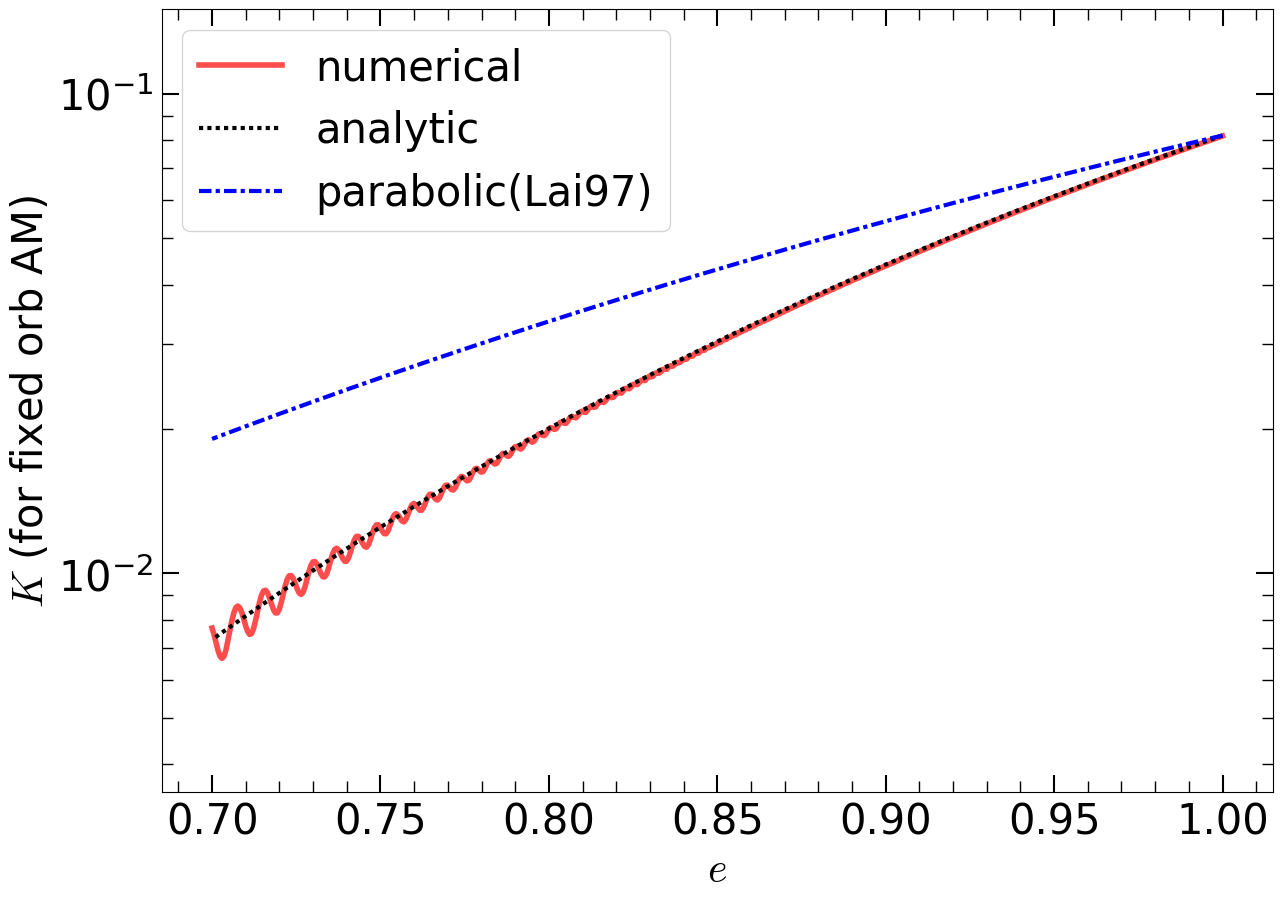}
    \vspace{-0.6cm}
    \caption{Tidal overlap integral $K$ computed using three different methods: direct numerical integration (red solid line), analytic result (eq.~\ref{eq:K_integral_result}) from saddle-point method (black dotted line), and the \citet{Lai1997} approximation for parabolic orbit (eq.~\ref{eq:K_integral_Lai97}) (blue dash-dotted line). We consider different orbital eccentricities but fix the specific orbital angular momentum such that $r_{\rm p}(1+e) = 2r_{\rm p,0}$ is conserved. The initial orbit is parabolic with pericenter radius $r_{\rm p,0}$ such that $\lambda_{\rm 0} = 2\omega/\Omega_{\rm p,0} = 10$, where $\Omega_{\rm p,0} = \sqrt{2GM_{\rm tot}/r_{\rm p,0}^3}$ is the initial pericenter angular frequency. As the eccentricity evolves, we have $\lambda(e) = 2\omega/\Omega_{\rm p} = 4(1+e)^2\lambda_0$ for a fixed specific orbital angular momentum.
    \vspace{0.6cm}
    }
    \label{fig:Ktidal_comparison}
\end{figure}

We further Taylor expand $\mathrm{e}^{i\lambda h_3 \tilde{z}^3}$ into polynomials
\begin{equation}
    \mathrm{e}^{i\lambda h_3 \tilde{z}^3} \approx 1 + i\lambda h_3 \tilde{z}^3,
\end{equation}
so the above integral $I(\lambda, \epsilon)$ has six terms (two for each $n=0, 1, 2$), and we ignore the higher-order ones and retain the following three terms
\begin{multline}
    I(\lambda, \epsilon)\approx \mathrm{e}^{-\lambda h_0} \int_{\mathcal{C}} \lrb{\sum_{n=0,1} g_n i^n \tilde{z}^{n-4} + g_0 \tilde{z}^{n-4} i\lambda h_3 \tilde{z}^3}\\\times\mathrm{e}^{-\lambda h_2 \tilde{z}^2} \d z.
\end{multline}
We skip the details and write down the results each of the three terms
\begin{equation}
    \int_{\mathcal{C}} g_0 \tilde{z}^{n-4} \mathrm{e}^{-\lambda h_2 \tilde{z}^2} \d z = {4\sqrt{\pi}\over 3} g_0 \lrb{\lambda h_2}^{3/2},
\end{equation}
\begin{equation}
    \int_{\mathcal{C}} i g_1 \tilde{z}^{-2} \mathrm{e}^{-\lambda h_2 \tilde{z}^2} \d z = \pi g_1 \lambda h_2,
\end{equation}
\begin{equation}
    \int_{\mathcal{C}} g_0 \tilde{z}^{n-4} i\lambda h_3  \tilde{z}^3 \mathrm{e}^{-\lambda h_2 \tilde{z}^2} \d z = -\pi g_0 \lambda h_3,
\end{equation}
so the integral $I(\lambda, \epsilon)$ is given by
\begin{equation}
    I(\lambda, \epsilon) = \mathrm{e}^{-\lambda h_0} \lrsb{{4\sqrt{\pi}\over 3} g_0 \lrb{\lambda h_2}^{3/2} + \pi g_1 \lambda h_2 - \pi g_0 \lambda h_3 }
\end{equation}
Putting in all the $g_n$ and $h_n$ factors, we obtain the final result for the tidal overlap integral
\begin{equation}\label{eq:K_integral_result}
\begin{split}
    K \approx {4W_{22}\over 3\sqrt{\pi}} & {\Omega_{*,1} \over \Omega_{p}} {\lambda^{3/2} \over (1-\epsilon)^2} \lrsb{1 - {\sqrt{\pi}\over 4}(1 +\epsilon)\lambda^{-1/2}} \times \\
    &\mathrm{exp}\lrsb{-{2\over 3}\lrb{1 + {2\over 5}\epsilon + {9\over 35}\epsilon^2 + {4\over 21}\epsilon^3}\lambda},
\end{split}
\end{equation}
where we find that $9\epsilon^2/35 + 4\epsilon^3/21$ are important as they are in the exponential argument but that the higher-order terms\footnote{Getting the next order $\mathcal{O}(\lambda^{-1})$ is quite involved, as one must include 4th order derivative in $h(z)$, the $n=2$ term in $g(z)$, and the 2nd order expansion of $\mathrm{e}^{i\lambda h_3 \tilde{z}^3}$ above. } $\mathcal{O}(\lambda^{-1})$ are negligible for sufficiently large $\lambda\gtrsim 10$. Our result (eq.~\ref{eq:K_integral_result}) is also in agreement with that of \citet{Lai1997} in the limit $\epsilon=0$,
\begin{equation}\label{eq:K_integral_Lai97}
    K(\epsilon=0) \approx {4W_{22}\over 3\sqrt{\pi}} {\Omega_{*,1} \over \Omega_{p,\rm PT}} \lambda_{\rm PT}^{3/2} \lrb{1 - {\sqrt{\pi}\over 4} \lambda_{\rm PT}^{-1/2}} \mathrm{e}^{-2\lambda_{\rm PT}/3},
\end{equation}
where the angular frequency at the pericenter $\Omega_{p,\rm PT}$ is for a parabolic orbit as considered by \citet{Press1977}
\begin{equation}
    \Omega_{p,\rm PT} = \sqrt{2G(M_{*,1}+M_{*,2})\over r_p^3}, \ \ \lambda_{\rm PT} \equiv {2\omega\over \Omega_{p,\rm PT}}.
\end{equation}
For an elliptical orbit with the same pericenter radius $r_p$, the physical pericenter frequency is
\begin{equation}
    \Omega_p = \sqrt{(1+e) G(M_{*,1}+M_{*,2})\over r_p^3} = \sqrt{1+e\over 2} \Omega_{p,\rm PT},
\end{equation}
and the mode frequency ratio is $\lambda = 2\omega/\Omega_p = \sqrt{2/(1+e)} \lambda_{\rm PT}$. 


\section{Second-order perturbation of the binary system}\label{sec:App_SA_perturb}
We provide below the expression of the second-order eccentricity perturbation of a binary system by the MBH using secular approximation in the parabolic orbit limit. Readers are referred to \citet{Hamers2019} for the derivation and more details. Here we only state the result for parabolic orbits in high perturber mass limit.

The inner eccentricity vector of the binary system prior to the perturbation is
\begin{align}
    \bs{e}_b & =e_x\hat{\bs{x}}+e_y\hat{\bs{y}}+e_z\hat{\bs{z}}\\
    & = \frac{1}{GM_b}\dot{\bs{r}}_b\times\lrb{{\bs{r}}_b\times\dot{\bs{r}}_b}-\hat{\bs{r}}_b,
\end{align}
where $\bs{r}_b$ is the separation between the binary. The normalized angular momentum vector $\bs{j}_b=j_x\hat{\bs{x}}+j_y\hat{\bs{y}}+j_z\hat{\bs{z}}$ is in the direction ${\bs{r}}_b\times\dot{\bs{r}}_b$ and has magnitude $\sqrt{1-e_b^2}$. The scale of eccentricity perturbation is controlled by the parameter $\epsilon_{\text{SA}}$:
\begin{equation}
    \epsilon_{\text{SA}} =\lrsb{\frac{1}{8}\frac{\MBH}{M_b}\lrb{\frac{a_b}{r_p}}^3}^{1/2}= \frac{1}{2\sqrt{2}}\lrb{\frac{r_t}{r_p}}^{3/2}.
\end{equation}

For the outer orbit in xy plane and outer pericenter in -x direction, the total perturbation to ${\bf e}_b,{\bf j}_b$ are
\begin{align}
    \Delta \bs{e}_b & = \epsilon_{\text{SA}}\bs{f}_e+\epsilon^2_{\text{SA}}\bs{g}_e,\\
    \Delta \bs{j}_b & = \epsilon_{\text{SA}}\bs{f}_j+\epsilon^2_{\text{SA}}\bs{g}_j,
\end{align}
where
\begin{align}
    \bs{f}_e = & -\frac{3\pi}{2}\lrb{3e_zj_y+e_yj_z}\hat{\bs{x}}+\frac{3\pi}{2}\lrb{3e_zj_x+e_xj_z}\hat{\bs{y}}\nonumber\\
                & +3\pi\lrb{e_yj_x-e_xj_y}\hat{\bs{z}},\\
    \bs{f}_j = & -\frac{3\pi}{2}\lrb{5e_ye_z-j_yj_z}\hat{\bs{x}}+\frac{3\pi}{2}\lrb{5e_xe_z-j_xj_z}\hat{\bs{y}},\\
    \bs{g}_e = & \frac{3}{16} \pi \Big[75 e_x^2 e_y-6 \pi \lrb{e_x \lrb{15 e_z^2-6 j_y^2+j_z^2}+6 e_y j_x j_y}\nonumber\\
                & +50 e_y^3+5 e_y \left(10 e_z^2+j_x^2-10 \left(j_y^2+2 j_z^2\right)\right)+50 e_z j_y j_z\Big] \hat{\bs{x}} \nonumber\\
                &- \frac{3}{16} \pi \Big[75 e_x^3+e_x \left(50 e_y^2+5 j_x^2+36 \pi  j_x j_y-150 j_z^2\right)\nonumber\\
                &+6 \pi  e_y \left(15 e_z^2-6 j_x^2+j_z^2\right)-10 j_x (5 e_y j_y+e_z j_z)\Big] \hat{\bs{y}} \nonumber\\
                &+\frac{3}{8} \pi \Big[-6 \pi  e_z \left(5 e_x^2+5 e_y^2-3 \left(j_x^2+j_y^2\right)\right)\nonumber\\
                &-5 (5 e_x e_y e_z+15 e_x j_y j_z-9 e_y j_x j_z+5 e_z j_x j_y)\nonumber\\
                &+12 \pi  j_z (e_x j_x+e_y j_y)\Big] \hat{\bs{z}}, \\
    \bs{g}_j = & \frac{3}{16} \pi \Big[75 e_x^2 j_y+60 \pi  e_x e_y j_y-6 \pi  j_x \left(10 e_y^2+15 e_z^2+j_z^2\right)\nonumber\\
                &-50 e_y e_z j_z+50 e_z^2 j_y+5 j_x^2 j_y\Big] \hat{\bs{x}} \nonumber\\
                &-\frac{3}{16} \pi \Big[15 e_x^2 (5 j_x+4 \pi  j_y)\nonumber\\
                &-10 e_x (6 \pi  e_y j_x+5 e_y j_y+15 e_z j_z)\nonumber\\
                &+5 j_x \left(10 e_y^2+j_x^2-2 j_z^2\right)+6 \pi  j_y \left(15 e_z^2+j_z^2\right)\Big] \hat{\bs{y}} \nonumber\\
                &-\frac{15}{8} \pi \left[5 e_x e_y j_z+5 e_x e_z j_y+5 e_y e_z j_x+j_x j_y j_z\right] \hat{\bs{z}}.
\end{align}
Note that in our models, the outer orbits do not necessarily lie on the xy plane. In those cases, proper coordinate rotations are first performed before applying the above perturbation results.

\section{Orbital circularization due to damping of dynamical tides}\label{sec:App-dyn-tides}
In the following we modify the results in \citet{Barker2020} based on physical arguments to roughly estimate the dynamical tides timescale of binaries $t_{a,\rm dyn}$ after the chaotic tides.

Consider the tidal dissipation in only one star of the binary of equal mass. The circularization timescale in a low-eccentricity limit is estimated to be \citep{Barker2020}
\begin{align}
    t_{e,\text{Barker}} & = \frac{2^{8/3}}{63\pi}Q^\prime\frac{P_{b}^{13/3}}{P_{\rm dyn}^{10/3}}, 
\end{align}
where $P_{\rm dyn}=2\pi\sqrt{R_*^3/(GM_*)}$ is the period associated with the dynamical time of the star, and $Q^\prime$ is the modified tidal quality factor. Here we consider the tidal damping from inertial waves, with the corresponding $Q^\prime$ \citep{Barker2020}:
\begin{align}
    Q^\prime & = Q_{\rm IW}^{\prime} \approx 10^7 \lrb{\frac{P_{\rm rot}}{10\,\text{d}}}^2,
\end{align}
where $P_{\rm rot}$ is the rotation period of the star.
For our binaries after the chaotic tides, there are two main differences. First, the high eccentricities of the binaries will limit the tidal interactions to close to the pericenter and reduce the overall efficiency of the circularization. We account for this effect by scaling $t_{e,\text{Barker}}$ with the ratio of the orbital period to the time near the pericenter:
\begin{align}
    t_{a,\rm dyn} = t_{e,\text{Barker}}\lrb{\frac{a_b}{r_{p,b}}}^{3/2}.
\end{align}
Second, while stars in a small-eccentricity binary tend to reach the synchronous rotation with the orbit, the high residual eccentricity in binaries after the chaotic tides will cause the star to spin near the orbital angular frequency at the pericenter. We therefore set $P_{\rm rot}$ to be
\begin{equation}
    P_{\rm rot} = 2\pi \sqrt{\frac{r_{p,b}^3}{GM_b(1+e_b)}}.
\end{equation}
Using binary conditions after chaotic tides (Figure~\ref{fig:ab-final}), we find that the dynamical tides timescale is less than one outer orbital period.
We emphasize that this extrapolation of $t_{a,\rm dyn}$ to highly eccentricities has huge uncertainties and is only meant to demonstrate the power of dynamical tides.

\end{appendix}

\end{document}